\documentclass[%
 aip,
 amsmath,amssymb,
 reprint,%
]{revtex4-1}

\usepackage{graphicx}
\usepackage{dcolumn}
\usepackage{bm}
\usepackage{adjustbox}
\usepackage[utf8]{inputenc}
\usepackage[T1]{fontenc}
\usepackage{mathptmx}
\usepackage{xcolor}
\usepackage{chemformula}

\usepackage{hyperref}
\hypersetup{
	colorlinks   = true,
	citecolor    = blue,
	linkcolor = blue,
    urlcolor = blue
}

\let\mrm\mathrm

\begin{document}

\preprint{AIP/123-QED}

\title{Recent progress in the JARVIS infrastructure for next-generation data-driven materials design}

\author{Daniel Wines}%
\affiliation{%
 Material Measurement Laboratory, National Institute of Standards and Technology,
Gaithersburg, Maryland 20899, USA 
}%

\author{Ramya Gurunathan}%
\affiliation{%
 Material Measurement Laboratory, National Institute of Standards and Technology,
Gaithersburg, Maryland 20899, USA 
}%

\author{Kevin F. Garrity}%

\affiliation{%
 Material Measurement Laboratory, National Institute of Standards and Technology,
Gaithersburg, Maryland 20899, USA 
}%

\author{Brian DeCost}%

\affiliation{%
 Material Measurement Laboratory, National Institute of Standards and Technology,
Gaithersburg, Maryland 20899, USA 
}%

\author{Adam J. Biacchi}%

\affiliation{%
 Physical Measurement Laboratory, National Institute of Standards and Technology,
Gaithersburg, Maryland 20899, USA 
}%

\author{Francesca Tavazza}%

\affiliation{%
 Material Measurement Laboratory, National Institute of Standards and Technology,
Gaithersburg, Maryland 20899, USA 
}%

\author{Kamal Choudhary}
 \email{kamal.choudhary@nist.gov}
 \affiliation{%
 Material Measurement Laboratory, National Institute of Standards and Technology,
Gaithersburg, Maryland 20899, USA 
}%

\date{\today}

\begin{abstract}
The Joint Automated Repository for Various Integrated Simulations (JARVIS) infrastructure at the National Institute of Standards and Technology (NIST) is a large-scale collection of curated datasets and tools with more than 80000 materials and millions of properties. JARVIS uses a combination of electronic structure, artificial intelligence (AI), advanced computation and experimental methods to accelerate materials design. Here we report some of the new features that were recently included in the infrastructure such as: 1) doubling the number of materials in the database since its first release, 2) including more accurate electronic structure methods such as Quantum Monte Carlo, 3) including graph neural network-based materials design, 4) development of unified force-field, 5) development of a universal tight-binding model, 6) addition of computer-vision tools for advanced microscopy applications, 7) development of a natural language processing tool for text-generation and analysis, 8) debuting a large-scale benchmarking endeavor, 9) including quantum computing algorithms for solids, 10) integrating several experimental datasets and 11) staging several community engagement and outreach events. New classes of materials, properties, and workflows added to the database include superconductors, two-dimensional (2D) magnets, magnetic topological materials, metal-organic frameworks, defects, and interface systems. The rich and reliable datasets, tools, documentation, and tutorials make JARVIS a unique platform for modern materials design. JARVIS ensures the openness of data and tools to enhance reproducibility and transparency and to promote a healthy and collaborative scientific environment.








\end{abstract}

\maketitle
\section{\label{sec:intro}Introduction}

\begin{figure}
\begin{center}
\includegraphics[width=10cm]{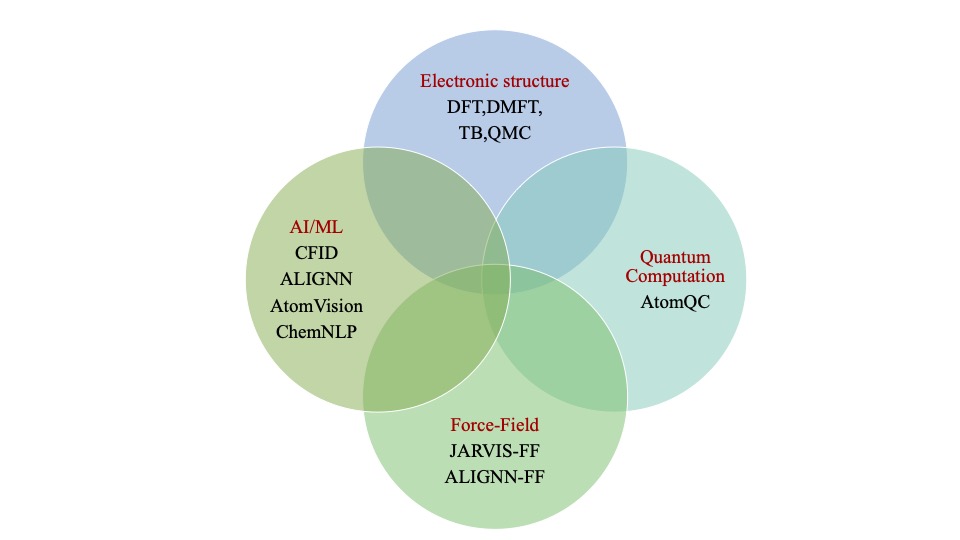}
\caption{Major areas of ongoing research as part of the JARVIS infrastructure.}
\label{areas}
\end{center}
\end{figure}

\begin{figure*}
\begin{center}
\includegraphics[width=0.95\textwidth]{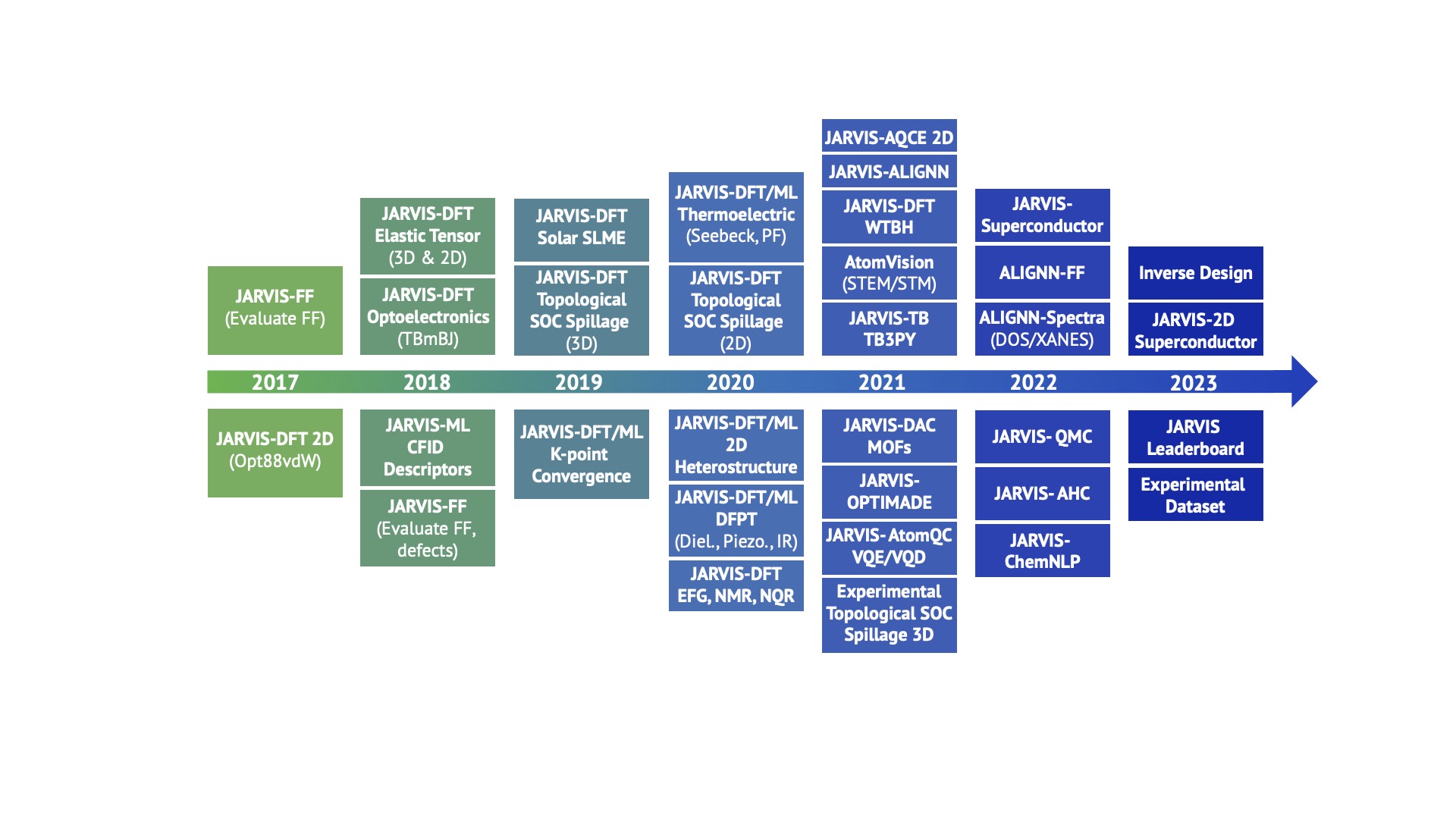}
\caption{An overview of the history of JARVIS-related projects since its creation in 2017 until present.}
\label{fig:history}
\end{center}
\end{figure*}

The Joint Automated Repository for Various Integrated Simulations (JARVIS)\cite{choudhary2020joint} is an integrated infrastructure to accelerate materials
discovery and design. The JARVIS infrastructure can be separated into electronic structure methods (density functional theory (DFT) \cite{PhysRev.136.B864}, tight binding \cite{Ashcroft}, dynamical mean field theory (DMFT) \cite{dmft}, many-body perturbation theory (GW) \cite{RevModPhys.74.601}, and Quantum Monte Carlo (QMC) \cite{RevModPhys.73.33}) \cite{martin2020electronic}, classical force-fields (FF) \cite{allen2017computer}, machine learning (ML) techniques \cite{hastie2009elements}, quantum computation algorithms \cite{nielsen2001quantum} and experiments \cite{leng2009materials}. JARVIS is motivated by the Materials Genome Initiative (MGI) \cite{warren2018materials} principles of developing open-access databases and tools to reduce the cost and development time of materials discovery, optimization, and deployment. A depiction of the major areas of ongoing research as part of the JARVIS infrastructure is depicted in Fig. \ref{areas} and the publicly available JARVIS tools are listed in Table \ref{jarvistools}.   

The main components of the JARVIS infrastructure (databases, user-friendly web applications, tools) are centrally located at \url{https://jarvis.nist.gov/}. The code behind the JARVIS infrastructure is located in a collection of separate repositories (i.e. JARVIS Tools, ALIGNN, ChemNLP, etc., see Table \ref{jarvistools}) that are centrally located in the main NIST GitHub repository (\url{https://github.com/usnistgov/}). Each of these repositories contains a myriad of code with specific separate installation instructions (i.e. using conda environments). For example, JARVIS Tools is a software package which contains a plethora of \textsc{python} functions and classes ($\approx$ tens of thousands of lines of code) used for automated materials simulations, post-processing of calculated data, and dissemination of results. Instructions for basic applications (i.e. example \textsc{python} code needed to screen materials in JARVIS or set up a basic DFT calculation) can be found in the JARVIS documentation (\url{https://pages.nist.gov/jarvis/}) or in the example \textsc{python} notebooks (see Section \ref{sec:notebooks}).


In the first three years since its creation in 2017 (see Fig. \ref{fig:history}), JARVIS-DFT grew to include standard material properties \cite{choudhary2020joint} such as formation energies, band gaps, elastic constants, piezoelectric
constants, dielectric constants, and magnetic moments, as well as more exotic properties such as exfoliation
energies for van der Waals (vdW) bonded materials \cite{choudhary2017high}, spin-orbit
coupling (SOC) spillage \cite{sos-1-jarvis,sos-2-jarvis,PhysRevB.103.155131}, improved meta-GGA band gaps \cite{choudhary2018computational},
frequency-dependent dielectric functions \cite{choudhary2018computational}, solar cell efficiency \cite{choudhary2019accelerated}, thermoelectric properties \cite{Choudhary_2020}, and Wannier
tight-binding Hamiltonians \cite{garrity2021database,garrity2021fast}. Protocols such as
automatic k-point convergence \cite{choudhary2019convergence} were developed to improve data reliability. JARVIS force field (JARVIS-FF) \cite{Choudhary_2018} offers a framework to use classical force fields to compute
material properties such as defect formation energies, bulk modulus,
and phonon spectra that can be utilized for molecular dynamics runs. Classical Force-field Inspired
Descriptors (CFID) \cite{choudhary2018machine} were introduced in 2018 as a part of JARVIS-ML. CFIDs represent the relation between the chemistry, structure and charge of a given material. By training CFIDs on JARVIS-DFT data, several classification and regression models have been developed. These include models to predict properties such as band gaps, formation energies, exfoliation energies, magnetic moments, thermoelectric properties, and several other properties \cite{choudhary2020joint}.

\begin{table*}[]
\caption{\label{jarvistools} A summary of the publicly available JARVIS tools.}
\begin{tabular}{l|l|l}

 Model Name & Link & Ref. \\
\hline
JARVIS Tools & \url{https://github.com/usnistgov/jarvis} & \cite{choudhary2020joint}\\
\hline
TB3PY & \url{https://github.com/usnistgov/tb3py} & \cite{garrity2021database}, \cite{garrity2021fast} \\
\hline
ALIGNN & \url{https://github.com/usnistgov/alignn} & \cite{Choudhary2021_ALIGNN}\\
\hline
AtomVision & \url{https://github.com/usnistgov/atomvision} & \cite{Choudhary2023_AtomVision} \\
\hline
ChemNLP & \url{https://github.com/usnistgov/chemnlp} & \cite{ChemNLP_arxiv}\\
\hline
AtomQC & \url{https://github.com/usnistgov/atomqc} & \cite{ChoudharyAtomQC} \\
\hline
JARVIS Notebooks & \url{https://github.com/JARVIS-Materials-Design/jarvis-tools-notebooks} & \cite{JarvisNbGitHub}\\
\hline
JARVIS Leaderboard & \url{https://github.com/usnistgov/jarvis_leaderboard} & \cite{choudhary2023large} \\
\hline
\end{tabular}
\end{table*}

In this review article, we will give an overview of the several major updates that have been made to the JARVIS infrastructure (see Fig. \ref{fig:history}). Recent updates to JARVIS-DFT, which now contains over 80000 materials, include identifying the anomalous quantum confinement effect in materials \cite{PhysRevMaterials.5.054602}, screening bulk magnetic topological materials \cite{PhysRevB.103.155131}, and screening bulk and two-dimensional (2D) superconducting materials \cite{jarvis-supercond-bulk,2dsc}. With regards to other electronic structure methods, tight binding models \cite{garrity2021database,garrity2021fast} and QMC methods \cite{RevModPhys.73.33,wines-crx3,wines2023quantum} have recently been added to the JARVIS infrastructure. JARVIS-ML has been expanded to include the Atomistic Line Graph Neural Network (ALIGNN) \cite{Choudhary2021_ALIGNN} model that has been utilized for fast and accurate property and spectra prediction of formation energies, band gaps, electron and phonon density-of-states \cite{Kaundiya2022_ALIGNN_eDOS,Gurunathan2023_ALIGNNphonons}, properties of metal-organic frameworks for carbon capture \cite{Choudhary2022_mof}, defect properties \cite{10.1063/5.0135382}, and properties of superconductors \cite{jarvis-supercond-bulk}. The ALIGNN model has also been recently used to develop universal force fields for the periodic table (ALIGNN-FF)  \cite{choudhary2023unified}. The AtomVision \cite{Choudhary2023_AtomVision} model has been added to JARVIS-ML, with the intention of generating and analyzing scanning tunneling microscope (STM) and high angle annular dark field (HAADF) scanning transmission electron microscope (STEM) images to accelerate the interpretation of experimental images. A natural language processing-based library for materials chemistry text data (ChemNLP) \cite{ChemNLP_arxiv} and tools to perform quantum computation algorithms \cite{ChoudharyAtomQC} such as variational quantum eigen solver (VQE) \cite{vqe} and variational quantum deflation (VQD) \cite{Higgott2019variationalquantum} have also been added to the JARVIS infrastructure. Finally, several experimental measurements have been performed to validate our computational predictions. In addition to the major recent updates of JARVIS, we will detail large-scale data efforts, educational notebooks, leaderboard, and external outreach.


\section{\label{sec:results}Electronic Structure}

\subsection{\label{sec:dft}Density Functional Theory}

\subsubsection{Magnetic Topological Materials Screening}\label{topscreen}

\begin{figure*}
\begin{center}
\includegraphics[width=0.8\textwidth]{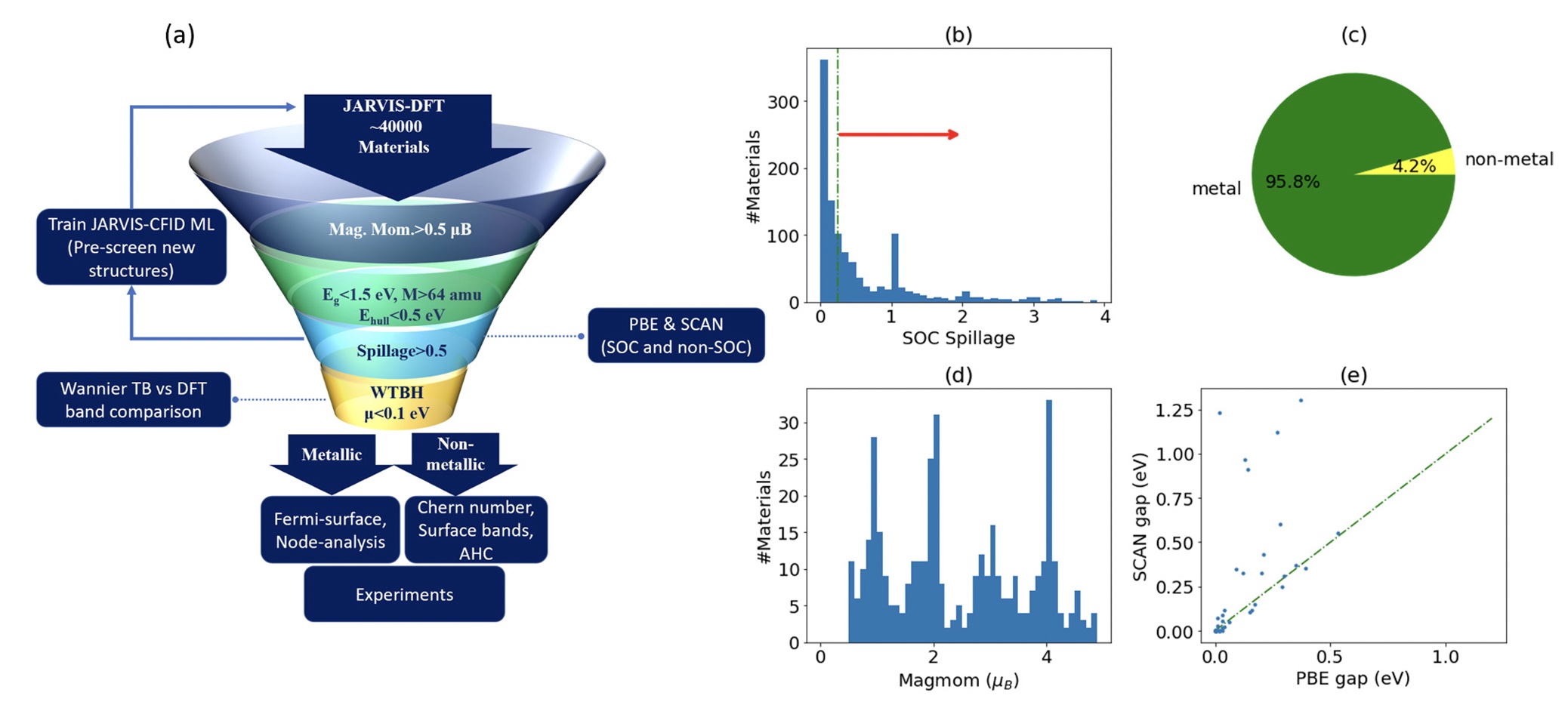}
\caption{(a) Flowchart depicting the screening process for high-spillage materials, (b) distribution of the spillage for all materials, (c) pie chart displaying high-spillage metals and insulators, (d) distribution of the magnetic moment for high-spillage structures, (e) band gaps computed with Perdew-Burke-Ernzerhof (PBE) \cite{PhysRevLett.77.3865} vs. Strongly Constrained and Appropriately Normed (SCAN) functionals \cite{PhysRevLett.115.036402}. Reproduced with permission from Phys. Rev. B 103, 155131 (2021). Copyright 2021 American Physical Society.}
\label{fig:spillage}
\end{center}
\end{figure*}

There have been few high-quality magnetic topological insulator and semimetal candidates identified in the literature, which can have potential applications in spintronics and quantum computation. We used a screening criteria based on spin-orbit spillage (SOS), which is a way to quantify spin-orbit-induced band inversion (a property of topological materials) by comparing the wave functions with and without spin-orbit coupling (SOC) \cite{PhysRevB.90.125133,sos-1-jarvis,sos-2-jarvis}. This study is an extension of previous work, which used SOS to screen for bulk nonmagnetic materials and magnetic and nonmagnetic 2D materials \cite{sos-1-jarvis,sos-2-jarvis}.

\begin{table}[hbt!]
\caption{A summary of magnetic topological materials: chemical formula (Form.), spacegroup number (Spg), JARVIS-DFT ID (JID) and maximum spillage values. Reproduced with permission from Phys. Rev. B 103, 155131 (2021). Copyright 2021 American Physical Society. \label{spillagetable}}
\begin{tabular}{llll}
\toprule
Form.  & Spg & JID & Spillage \\
\toprule
Mn$_2$Sb    & $P6_3$/$mmc$     & 15693                         & 0.5                          \\
NaMnTe$_2$  & $P\bar{3}m1$       & 16806                         & 1.04                         \\
Rb$_3$Ga    & $Fm\bar{3}m$      & 38248                         & 0.47                         \\
CoSI     & $F\bar{4}3m$       & 78508                         & 0.69                         \\
Mn$_3$Sn    & $P6_3$/$mmc$     & 18209                         & 0.79                         \\
Sc$_3$In    & $P6_3$/$mmc$     & 17478                         & 1.01                         \\
Sr$_3$Cr    & $Pm\bar{3}m$       & 37600                         & 1.01                         \\
Mn$_3$Ge    & $Fm\bar{3}m$       & 78840                         & 3.01                         \\
NaRuO$_2$   & $R\bar{3}m$        & 8122                          & 0.5                          \\
CoNb$_3$S$_6$  & $P6_322$       & 21459                         & 1.03                         \\
Y$_3$Sn     & $P6_3$/$mmc$     & 37701                         & 0.29                         \\
CaMn$_2$Bi$_2$ & $P\bar{3}m1$       & 18532                         & 1.17 \\
\toprule
\end{tabular}
\end{table}

We used systematic high-throughput DFT calculations to identify magnetic topological materials from the over 40000 bulk materials in the JARVIS-DFT database \cite{PhysRevB.103.155131}. First, we screen materials with net magnetic moment > 0.5 $\mu_{\textrm{B}}$ and SOS > 0.25, resulting in 25 insulating and 564 metallic candidates. We then perform Wannier tight-binding Hamiltonian (WTBH)-based techniques to calculate Wannier charge centers, Chern numbers, anomalous Hall conductivities (AHC), surface band structures, and Fermi surfaces to determine interesting topological characteristics of the screened compounds. After narrowing down the search, we experimentally synthesized and characterized a few candidate materials such as CoNb$_3$S$_6$ and Mn$_3$Ge. 

The full workflow is given in Fig. \ref{fig:spillage}a), while a full analysis of the data trends for the materials is given in Fig. \ref{fig:spillage}b) - e). A summary of candidate materials with high values of SOS is given in Table \ref{spillagetable}. Further analysis of the electronic band structure (with and without SOC) and k-dependent spin-orbit spillage was conducted. Strong focus was placed on Y$_3$Sn (JVASP-37701), which is a candidate semimetal, and further analysis of the Fermi surface (001) surface band structure, nodal points-lines, and AHC was performed. In addition, strong focus was placed on NaRuO$_2$ (JVASP-8122), which is a candidate Chern insulator, and further analysis of the Wannier charge center and AHC were performed. Further details of computational screening, methodologies, and specific calculated results can be found in Ref. \onlinecite{PhysRevB.103.155131}.

\subsubsection{Anomalous Quantum Confinement Effect}

\begin{figure}
\begin{center}
\includegraphics[width=0.4\textwidth]{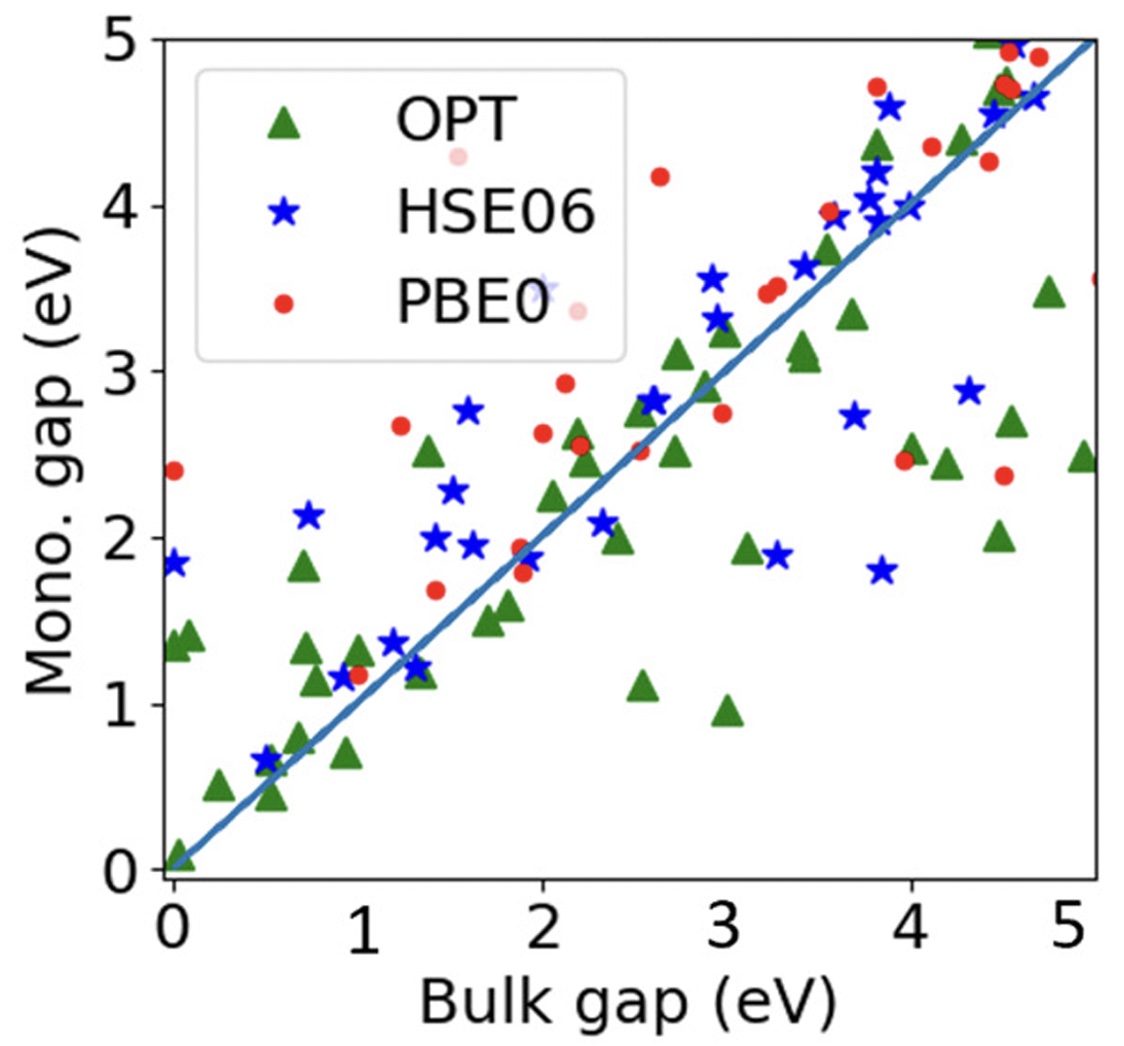}
\caption{Bulk vs. monolayer band gaps using OptB88vdW (OPT), HSE06 and PBE0, demonstrating the AQCE. Reproduced with permission from Phys. Rev. Mater. 5, 054602 (2021). Copyright 2021 American Physical Society.} 

\label{fig:AQCE}
\end{center}
\end{figure}

Quantum confinement effects, where the electronic band gap of a bulk material is lower in magnitude than the band gap of its 2D counterpart, are prevalent for vdW bonded materials. In contrast, it is possible that this band gap trend is reversed, resulting in an anomalous quantum confinement effect (AQCE). We calculated the band gaps for bulk and corresponding 2D counterparts using DFT, starting from structures in the JARVIS-DFT database. We used semilocal functionals (OptB88vdW \cite{klimevs2009chemical}) for $\approx$1000 materials and hybrid functionals (HSE06 \cite{doi:10.1063/1.1564060} and PBE0 \cite{doi:10.1063/1.478522}) for $\approx$50 materials. We identify 65 AQCE candidates with OptB88vdW, but only confirm this peculiar effect with hybrid functionals for 14 materials. Depending on the material system, the band gap differences (between bulk and monolayer) can range from less than 0.5 eV to 2 eV. Fig. \ref{fig:AQCE} depicts these computed results. A large portion of the AQCE candidates are hydroxides and oxide hydroxides (AlOH$_2$, Mg(OH)$_2$, Mg$_2$H$_2$O$_3$, Ni(OH)$_2$, SrH$_2$O$_3$), alkali-chalcogenides (RbLiS and RbLiSe), and Sb-halogen-chalcogenides (SbSBr, SbSeI).


Strikingly, we found examples of 0D and 1D structures included in the 14 AQCE candidates. To quantify the effect of SOC on the band structure predictions and determine the SOS (similar procedure to Section \ref{topscreen}) to screen for topological properties, we performed Perdew-Burke-Ernzerhof (PBE)-based SOC calculations. From these results, we found that SOC does not significantly alter the band gap and none of the 14 materials have topological properties. We further investigated the change in electronic structure and bond distances with the goal of understanding the AQCE. We found that in ACQE materials, there is a lowering of the conduction band in the 2D structures with changes in the contribution of the p$_z$ orbitals (z is the non-periodic direction). We also find for structures that contain OH, there are significant changes in the H-H bond distances, which can be responsible for the AQCE. More details on this work and a full list of AQCE materials can be found in Ref. \onlinecite{PhysRevMaterials.5.054602}.

\subsubsection{Bulk and 2D BCS Superconductors}\label{Bulk and 2D BCS Superconductors}

\begin{table}[hbt!]
\caption{JARVIS screening workflow for some of the potential candidate superconductors: ($T_c$), chemical formula (Form.), spacegroup number (Spg), JARVIS ID (JID), Inorganic Crystal Structure Database ID (ICSD) \cite{belsky2002new} wherever available, JARVIS-DFT based formation energy ($E_{form}$ (eV/atom)) and energy above \textcolor{black}{convex hull} ($E_{hull}$ (eV)). Reproduced with permission from npj Computational Materials 8, 244 (2022). Copyright 2022 Nature Publications.\label{tab:results}}
\begin{tabular}{@{}lllllll@{}}
\toprule
Form. & Spg & JID&ICSD&$E_{form}$&$E_{hull}$&$T_C$(K)\\
\toprule
MoN & 187 & 16897&187185   &-0.47 &0.09&33.4  \\
CaB$_2$ & 191 & 36379  &237011 & -0.25&0.09&31.0  \\
ZrN&194&13861&161885&-1.76&0.18&30.0\\
VC & 225 & 19657&619079   &-0.48&0.06&28.1 \\
V$_2$CN & 123 & 105356&-   &-0.82&0.11&26.2 \\
Mn &225&25344&41509&0.08&0.08&23.0\\
NbFeB&187&4546&-&-0.15&0.39&22.1\\
NbVC$_2$&5&102190&-&-0.46&0.08&21.9\\
ScN &225&15086&290470 &-2.15&0.0&20.8\\
LaN$_2$&2&118592&&-1.05&0.0&20.4\\
VRu&221&19694&106010 &-0.22&0.01&20.3\\
TiReN$_3$&161&36745&-&-0.68&0.10&20.0\\
B$_2$CN&51&91700&183794&-0.53&0.19&19.4\\
KB$_6$ &221 &20067 &98987& -0.09 &0.0&19.0\\
ZrMoC$_2$&166&99893&-&-0.49&0.08&17.9\\
TaB$_2$ &191&20082&30420&-0.60&0.0&17.2\\
NbS&194&18923&44992& -0.98&0.05&17.0\\
TaVC$_2$&166&101106&-&-0.54&0.05&16.3\\
TaC&187&36405&&-0.24&0.40&16.1\\
MgBH&11&120827&-&-0.03&0.11&15.5\\
CoN&216&14724&236792 &-0.02&0.0&15.0\\
NbRu$_3$C&221&8528&77216&-0.02&0.19&15.0\\

\toprule
\end{tabular}
\end{table}

The search for superconducting materials with high transition temperatures ($T_C$) has been a goal of condensed matter physicists \cite{poole2013superconductivity,rogalla2011100} since the discovery of superconductivity in 1911 \cite{kamerlingh1911resistance}. The search for novel superconductors can be expedited with more data-driven and systematic approaches. In order to identify high-$T_C$ conventional Bardeen–Cooper–Schrieffer (BCS) superconductors \cite{cooper2010bcs, giustino2017electron}, a curated database of materials that can assist in screening candidates and an efficient high-throughput workflow to perform electron-phonon coupling (EPC) calculations are both required. The EPC that can be used to reliably predict $T_C$ can be obtained from density functional theory perturbation theory (DFT-PT) calculations \cite{giustino2017electron, PhysRevB.101.134511}.

\begin{figure*}
\begin{center}
\includegraphics[width=0.8\textwidth]{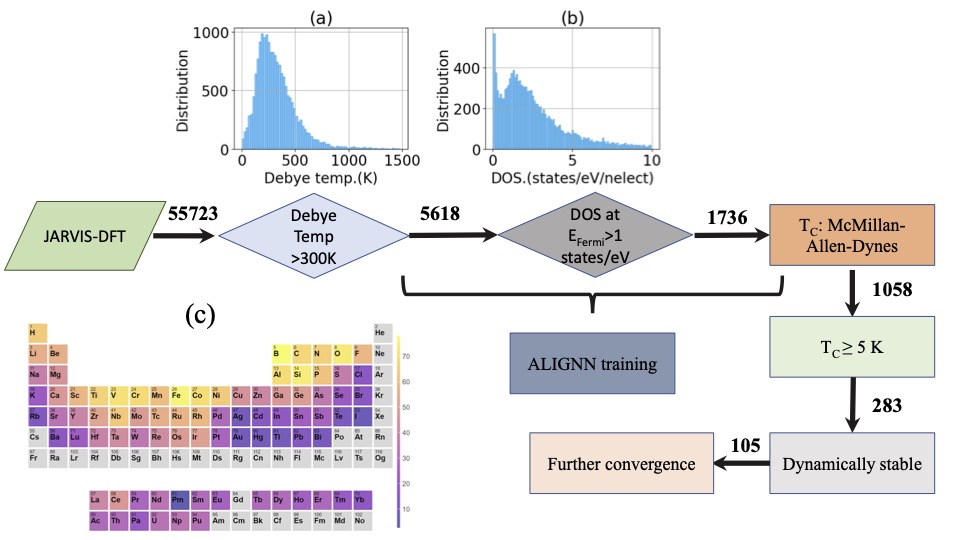}
\caption{The main workflow used to identify bulk BCS superconductors. a) statistical distribution of the Debye temperature (in K) and b) statistical distribution of the electronic density of states (DOS) (in states per eV per number of electrons) at the Fermi level, c) the likelihood that a material contains a given element for $\theta_D$ greater than 300 K. Reproduced with permission from npj Computational Materials 8, 244 (2022). Copyright 2022 Nature Publications.}
\label{sc-chart}
\end{center}
\end{figure*}

We combined several approaches, each with different levels of computational expense, to design a high-throughput workflow to discover new BCS superconductors. This workflow, depicted in Fig. \ref{sc-chart}, begins with a pre-screening step to identify materials in the JARVIS-DFT database with a high electron density of states (DOS) at Fermi-level ($N(0)$) and high Debye temperature ($\theta_D$). Next, we developed and applied a DFT-PT workflow to obtain the EPC properties and calculated $T_C$ using the McMillan-Allen-Dynes formula \cite{mcmillan1968transition} (with initially low k-point and q-point convergence settings). Before applying this DFT-PT workflow to materials from the pre-screening step, we validated our methods by benchmarking the workflow for several well-known superconductors. We performed additional k-point and q-point convergence for the top candidates from our pre-screening step. As discussed in Section \ref{sec:alignn}, we used our EPC computed data to develop a deep-learning property prediction model for superconducting properties using the the atomistic line-graph graph neural network (ALIGNN).

We specifically pre-screened 1736 materials with high Debye temperature and electronic density of states. From our pre-screening step, we identified 1736 candidates (high electronic density of states at the Fermi level and high Debye temperature) and performed DFT-PT for 1058 of them to obtain EPC properties and $T_C$. From this, we found 105 stable structures with a $T_C$ above 5 K (top candidates are shown in Table \ref{tab:results}). The superconductors with the highest $T_C$ include MoN, VC, Mn, MnN, LaN$_2$, KB$_6$, and TaC. Most notably, we discover a new hexagonal form of MoN, which has not been experimentally observed (in contrast to the superconducting rock-salt phase which has a $T_C$ of 30 K \cite{inumaru2008high,wang2015hardest}). Further details of this work on 3D BCS superconductors can be found in Ref. \onlinecite{jarvis-supercond-bulk}.

\begin{figure*}
\begin{center}
\includegraphics[width=0.7\textwidth]{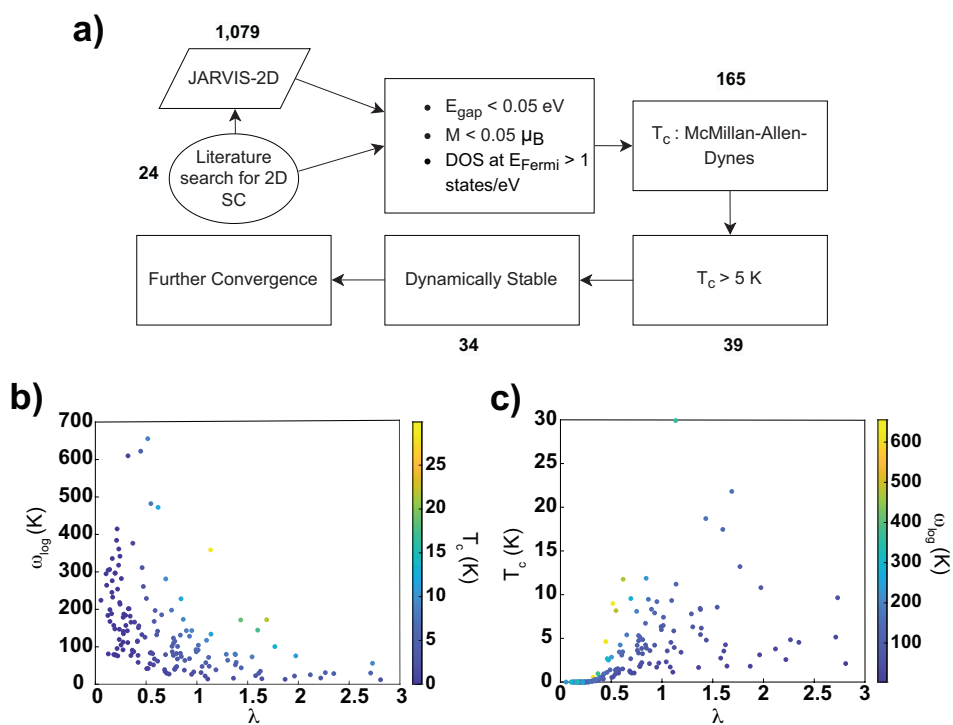}
\caption{a) High-throughput workflow used to screen for 2D superconductors with high $T_c$ and b) - c) the relationship between $\lambda$ and $\omega_{log}$ (electron-phonon coupling) for the materials in this study. Reproduced with permission from Nano Letters 23, 969-978 (2023). Copyright 2023 American Chemical Society.}
\label{sc-chart-2d}
\end{center}
\end{figure*}

Superconductivity in 2D has attracted attention \cite{doi:10.1021/acs.nanolett.0c05125,PhysRevB.96.094510,Mg2B4C2} due to the potential applications in quantum interferometers, superconducting transistors and superconducting qubits \cite{app1,PhysRevB.79.134530,doi:10.1126/science.299.5609.1045,doi:10.1063/1.3521262,app2}. Since very few high $T_C$ 2D materials have been computationally or experimentally identified, we decided to extend our high-throughput workflow to 2D superconductors. First, we pre-screened over 1000 2D materials in the JARVIS-DFT database on the basis of DOS at the Fermi level, electronic band gap, and the total magnetic moment. This screening criterion is modified from our workflow on bulk superconductors because the elastic tensor is available only for a limited number of monolayers in JARVIS (it is more computationally expensive to calculate for 2D structures). This modified screening procedure is based on the fact that a candidate 2D superconductor will have a high density of states at the Fermi level (metallic) and zero magnetic moment per unit cell. We additionally found 24 monolayers based on a literature search of bulk and monolayer superconductors. A full depiction of this workflow for 2D materials and a summary of the results are given in Fig. \ref{sc-chart-2d}.

Several nitrides, borides and carbides are found to be among the 2D materials found to have a high $T_C$. Also, many oxide and niobium-based structures and transition metal dichalcogenides (such as NbS$_2$ and NbSe$_2$) are found to be good candidate superconductors. Similar to a recent computational study \cite{doi:10.1021/acs.nanolett.0c05125}, we find W$_2$N$_3$ to possess a significantly high $T_c$ of 18.7 K. We observe the highest $T_c$ of 21.8 K for 2D Mg$_2$B$_4$N$_2$, which has been previously undiscovered in 2D and 3D form. In addition, we studied 2D analogs of non-layered materials such as ScC, NbC, B$_2$N and MgB$_2$ and oxide-based materials such as TiClO, ZrBrO and NbO$_2$, all of which are superconducting. Further information on 2D superconductors can be found in Ref. \onlinecite{2dsc}.

\subsection{\label{sec:tb}Tight Binding}

There are two types of tight-binding projects available in JARVIS: 1) Wannier tight-binding Hamiltonians (WTBH) \cite{garrity2021database}, 2) a parametrized universal tight-binding model fit to first principles calculations (ThreeBodyTB.jl) \cite{garrity2021fast}. The WTBH database provides a computationally efficient way to interpolate and understand the electronic properties of a set of 1771 preselected materials, based on a DFT calculation for each of those materials. The quality of the WTBH is evaluated by comparing the Wannier band structures to directly calculated DFT band structures including SOC. The WTBH database is used for predicting the AHC, surface band structures, and various topological indexes. 

In contrast to the WTBH database, the goal of the ThreeBodyTB.jl parametrized tight-binding model is to produce a tight-binding Hamiltonian and total energy \textit{without} doing a computationally expensive DFT calculation first. Because tight-binding uses a minimal basis set of atomic orbitals, the calculations are up to three orders of magnitude faster than comparable plane-wave DFT calculations, enabling computationally efficient materials prediction. Despite their simplicity, tight-binding approaches incorporate single-particle quantum mechanics as well as electrostatics as a self-consistency step\cite{elstner1998self,frauenheim2000self,koskinen2009density}. This built-in physics can enable improved predictions outside the set of training data, relative to classical force-fields or pure machine-learning approaches. 

Unlike typical parametrized tight-binding models that consider only interactions between pairs of atoms when generating the tight-binding Hamiltonian\cite{koskinen2009density, hourahine2020dftb+}, our model includes three-body contributions that modify the two-body contributions as well. These extra terms allow for improved transferability as compared to simpler models, at the cost of needing to fit more parameters. 

\begin{figure}
\begin{center}
\includegraphics[width=3.4in]{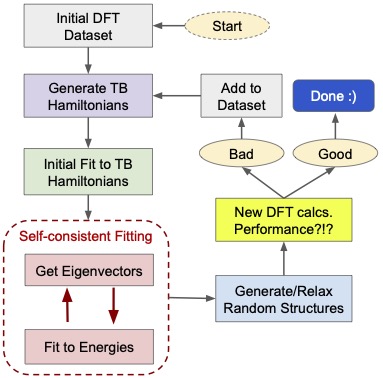}
\caption{Overview of the three-body tight binding (TB) model fitting workflow. Reproduced with permission from Phys. Rev. Mater. 7, 044603  (2023). Copyright 2023 American Physical Society.}
\label{fig:tb3_workflow}
\end{center}
\end{figure}

Our fitting procedure is summarized in Fig. \ref{fig:tb3_workflow}. For a given elemental or binary system, we first generate a set of standard crystal structures, perform DFT calculations, and fit an initial parameter set to reproduce the band structures and total energies. Then, we employ an active learning strategy to test and improve the model by using the current model to relax randomly generated crystal structures \cite{Pickard_2011} and test our tight-binding results versus new DFT calculations. If the results are poor, we add these new structures to our fitting database and repeat the process until the results improve.

\begin{figure}
\begin{center}
\includegraphics[width=3.4in]{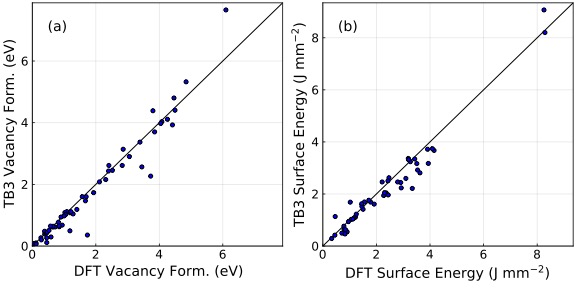}
\caption{DFT results vs. three-body tight-binding results for unrelaxed a) point vacancy formation energy (in eV) and b) $(111)$ surface energies (in $J$ $mm^{-2}$) of various elemental solids. The results from tight-binding are out-of-sample. Reproduced with permission from Phys. Rev. Mater. 7, 044603  (2023). Copyright 2023 American Physical Society.}
\label{fig:tb3_defect_surface}
\end{center}
\end{figure}

Our current parameter set can predict total energies, volumes, and band gaps with comparable accuracy to machine learning approaches, as well as produce band structures. Importantly the results generalize to surfaces and vacancy calculations that are completely outside the fitting dataset, as shown in Fig. \ref{fig:tb3_defect_surface}. For testing results and details see Ref. \onlinecite{garrity2021fast}. The \textsc{julia} code with a \textsc{python} interface is available, and an underlying DFT database with over one million materials is available in JARVIS-QETB.

\subsection{\label{sec:qmc}Quantum Monte Carlo}

A recent effort of the JARVIS infrastructure has been to incorporate many-body methods that go beyond the standard accuracy of DFT for selected materials that have a complicated or correlated electronic structure. 
Diffusion Monte Carlo (DMC) \cite{RevModPhys.73.33} is a many-body correlated electronic structure method that has been applied successfully to the calculation of electronic and magnetic properties of a variety of periodic systems. It involves solving the imaginary-time Schr\"{o}dinger equation for the near-exact ground state wavefunction using projector techniques (more details can be found in Ref. \onlinecite{RevModPhys.73.33}). Although it is a more computationally expensive method, DMC has a weaker dependence on the starting density functional and Hubbard U parameters \cite{PhysRevB.57.1505}, scales similarly to DFT with respect to the number of electrons in the simulation ($\sim N^{3-4})$ \cite{RevModPhys.73.33}, and can achieve results that are more accurate than DFT \cite{RevModPhys.73.33}.

\subsubsection{Systematic Benchmark of 2D CrX$_3$}

\begin{figure*}
\begin{center}
\includegraphics[width=0.8\textwidth]{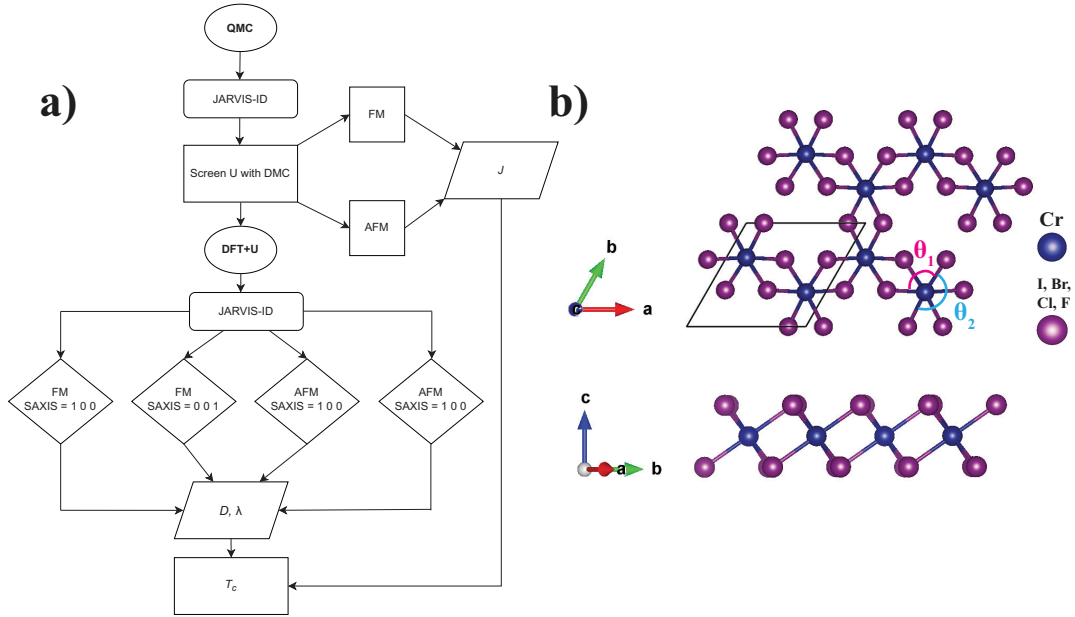}
\caption{a) High throughput workflow used to calculate the magnetic properties using DFT+U in conjunction with QMC for a 2D material and b) side and top views of the structure of 2D CrX$_3$ (X = I, Br, Cl, F). Reproduced with permission from J. Phys. Chem. C 127, 1176-1188 (2023). Copyright 2023 American Chemical Society.}
\label{fig:crx3}
\end{center}
\end{figure*}

We designed a workflow that applied a combination of DFT+U and DMC techniques to compute accurate magnetic properties for 2D CrX$_3$ materials (X = I, Br, Cl, F) \cite{wines-crx3}. We chose these materials (depicted in Fig. \ref{fig:crx3}b) as a case study since they have been experimentally synthesized \cite{cri3,crbr-exp,crcl3-exp}, have a nonzero critical temperature \cite{cri3,olsen-data}, and have been studied with DFT extensively \cite{olsen-data}. Our first-principles data can be mapped to a 2D model spin Hamiltonian to extract useful observable quantities such as $T_c$ \cite{Torelli_2018,Lado_2017}. In our case, this Hamiltonian was a function of Heisenberg isotropic exchange ($J$), easy axis single ion anisotropy ($D$) and anisotropic exchange ($\lambda$). Note, this $\lambda$ is different than the $\lambda$ that represents the electron-phonon coupling strength in superconductors (previously mentioned).      
To obtain $J$, $\lambda$ and $D$, we performed spin-orbit (noncollinear) DFT+U calculations by rotating the easy axis by 90$^{\circ}$ and calculating the energy difference between the rotated and non-rotated configurations for ferromagnetic (FM) and antiferromagnetic (AFM) separately. We automated these four calculations using the JARVIS workflow, where four distinct total energy values were obtained for each structure. We benchmarked this for 2D CrI$_3$ (JVASP-76195), CrBr$_3$ (JVASP-6088), CrCl$_3$ (JVASP-76498) and CrF$_3$ (JVASP-153105) using multiple DFT functionals and values of U. The influence of the geometric structure on the magnetic properties was also assessed (see Ref. \onlinecite{wines-crx3} for more details).

It is possible to systematically improve these results with QMC. This can be accomplished by variationally determining the optimal U value with DMC and computing a statistical bound for the $J$ parameter, which involves DMC calculations for the FM and AFM states separately. In comparison to the previous noncollinear (spin-orbit) DFT calculations, the energies in QMC are from collinear (spin-polarized) calculations. Due to the fact that spin-orbit calculations are limited in DMC at the moment, we are forced to neglect the $\lambda$ contribution (spin-orbit dependent term) when computing $J$ with DMC. However, since $J >> \lambda$, this has negligible impact on the end result for $J$. Fig. \ref{fig:crx3}a) displays the full QMC and DFT+U high-throughput workflow that allows us to accurately estimate the 2D critical temperature\cite{Torelli_2018} with the extracted $J$ from DMC and the anisotropy parameters ($D$, $\lambda$) from DFT+U (at the optimal U value determined from DMC). We estimated a maximum value of 43.56 K for the T$_c$ of CrI$_3$ and 20.78 K for the T$_c$ of CrBr$_3$. Additionally, we present a comparison between the magnetic moments and spin-density computed with DFT+U and DMC. For more details of this work see Ref. \onlinecite{wines-crx3}.

\begin{figure}
\begin{center}
\includegraphics[width=0.4\textwidth]{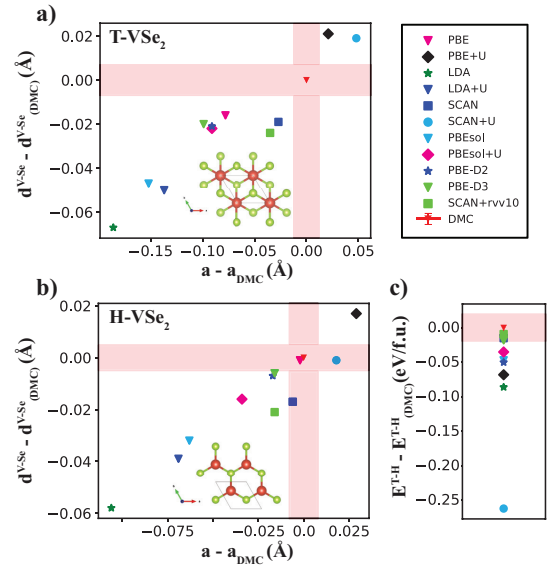}
\caption{Deviation of the structural properties (lattice constant ($a$) and V-Se distance ($d^{\textrm{V}-\textrm{Se}}$)) compared to the DMC computed structural properties for a) T-VSe$_2$ and b) H-VSe$_2$ and c) the deviation of T - H energy compared to the DMC computed T - H energy (E$^{\textrm{T} - \textrm{H}}$) for different DFT functionals (U = 2 eV), where the DMC error bar (standard error about the mean) is indicated by red bars. The side and top view of the geometric structure is shown in the insets. Reproduced with permission from J. Phys. Chem. Lett. 14, 3553-3560 (2023). Copyright 2023 American Chemical Society.}
\label{fig:vse2}
\end{center}
\end{figure}

\subsubsection{Structure and Phase Stability of 2D 1T- and 2H-VSe$_2$}

Throughout the theoretical and experimental landscape, there have been controversies involving 2D VSe$_2$ such as reports of near-room temperature ferromagnetism (Curie temperature ranging from 291 K to 470 K) \cite{vse2,https://doi.org/10.1002/adma.201903779,vse2-exp,vse2-bkt}. A coupling of structural parameters to magnetic properties is a likely cause for the discrepancies in experimental and calculated results \cite{vse2,https://doi.org/10.1002/adma.201903779,vse2-exp,vse2-bkt,vse2-moment-exp} for monolayer VSe$_2$ in the T (octahedral phase (1T)-centered honeycombs) phase and the H (the trigonal prismatic (2H)-hexagonal honeycombs) phase. These structures are shown in the insets of Fig. \ref{fig:vse2}. Both the T- and H-phases have a close lattice match and similar total energies, which makes it a challenge to distinguish which phase is being observed experimentally \cite{struc-phase,C9CP03726H,https://doi.org/10.1002/adma.201903779,vse2}. In order to resolve the discrepancies in geometric properties and relative phase stability of 2D VSe$_2$ (T- and H-phase), we used a combination of DFT, DMC and a newly developed surrogate Hessian line-search geometric optimization tool \cite{doi:10.1063/5.0079046}.

In Fig. \ref{fig:vse2}, DFT benchmarking results are shown for multiple DFT functionals (with and without U correction) for structural parameters such as lattice constant ($a$) and V-Se distance ($d^{\textrm{V}-\textrm{Se}}$) and the relative energy between the T- and H-phase (E$^{\textrm{T} - \textrm{H}}$). We observe a large deviation between DFT methods, which indicates the need to incporporate more accurate theories such as DMC. Using DMC, we computed the lattice constants to be 3.414(12) \AA\space and 3.335(8) \AA\space for T-VSe$_2$ and H-VSe$_2$ respectively. We also computed the V-Se distance to be 2.505(7) \AA\space and 2.503(5) \AA\space for T-VSe$_2$ and H-VSe$_2$. We find the DMC relative energy to be 0.06(2) eV per formula unit, which indicates that the H-phase is energetically more favorable than the T-phase in freestanding form. Using the DMC potential energy surface, we estimated a phase diagram between the phases and found that applying small amounts of strain can induce a phase transition. We also computed the magnetic moments and spin densities with DMC and find substantial differences between DMC and DFT+U. More detailed information on this study can be found in Ref. \onlinecite{wines2023quantum}.

\section{\label{sec:ai/ml}AI/ML}


\subsection{\label{sec:alignn}ALIGNN}

Non-Euclidean graphs are increasingly being used to represent crystal structures in deep-learning models. In comparison to composition-only based descriptors, the graph representation preserves the bond connectivity of atoms. Graph neural networks (GNN) are deep-learning frameworks which perform inference on graph data structures, and several high-performing GNN models have been proposed for the prediction of material properties, including but not limited to: SchNet\cite{Schutt2018SchNet}, Crystal Graph Convolutional Neural Networks (CGCNN)\cite{Xie2018CGCNN}, improved Crystal Graph Convolutional Neural Networks (iCGCNN)\cite{Park2020_iCGCNN}, MatErials Graph Network (MEGNet)\cite{Chen2019MEGNet}, and OrbNet\cite{Qiao2020}. In these frameworks, the graph nodes represent atoms and encode for elemental features, while the edges represent bonds and encode for bond distances. Therefore, only pairwise interactions are explicitly encoded in the materials representation. Through the use of multiple graph convolutional layers in the neural network, nodes (atoms) are updated based on neighboring states, allowing for an implicit handling of many-body interactions\cite{Choudhary2021_ALIGNN}. 

\begin{figure*}
\begin{center}
\includegraphics[width=0.8\textwidth]{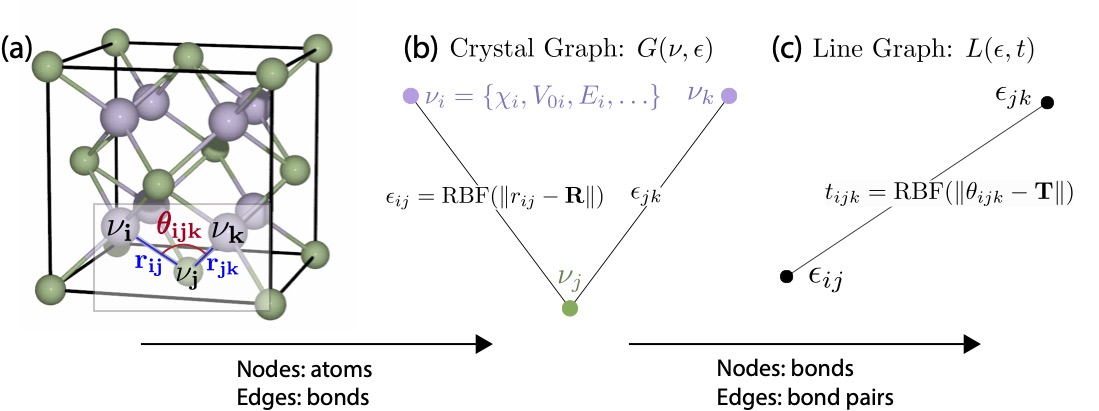}
\caption{a) A schematic of the encoding of the crystal and line graphs for \ch{Mg_2Si}. b) Depicts the crystal graph, where the nodes represent the atomic sites and include an atomic feature set (ionization energy ($E$), volume per atom ($V_{0i}$), electronegativity ($\chi$)). Bonds are represented by edges in the crystal graph and bond distances ($r_{ij}$) are represented by the edge features, which use a radial basis function (RBF) to encode $r_{ij}$ in the model. c) Depicts the line graph (constructed from the previous crystal graph), where the bonds of the structure are now represented by the nodes. Pairs of bonds with an atom in common (``triplets") featurized by the bond angles are represented by the edges, which are also encoded using a RBF. Reproduced with permission from npj Computational Materials 7, 185 (2021). Copyright 2021 Nature Publications}

\label{fig:alignn_schematic}
\end{center}
\end{figure*}

However, several material properties, for example those related to wave-like electronic and phononic states, are influenced by local geometric distortions captured by changes in bond angles. In order to explicitly encode bond angles and three-body configurations of atoms, we introduced the Atomistic Line Graph Neural Network (ALIGNN) \cite{Choudhary2021_ALIGNN} into the suite of JARVIS tools. Following the original work on line graph neural networks by Chen \textit{et al.}\cite{Chen2018LineGraph}, the ALIGNN framework sequentially updates two graph representations: 1) the crystal graph with nodes representing atoms and edges representing bonds, and 2) the line graph built from the crystal graph with nodes representing bonds and edges representing bond pairs sharing a common atom or triplets of atoms. Note that the edges of the crystal graph and nodes of the line graph share the same latent representation. Figure \ref{fig:alignn_schematic} depicts how the compositional and structural features of a material are encoded in both graph representations. The atomistic feature set ($\nu_i$ for the $i$\textsuperscript{th} atom) describing a crystal graph node includes the following elemental descriptors: electronegativity, covalent radius, group number, block, valence electron count, atomic volume, first ionization energy, and electron affinity\cite{Choudhary2021_ALIGNN}. The bond features ($\epsilon_{ij}$ for pairs of atoms $i$ and $j$) are the bond distances, represented using a radial basis function (RBF) expansion. Finally, the triplet features ($t_{ijk}$ for set of atoms $i$, $j$ and $k$) are an RBF expansion of the bond angle cosines.

The ALIGNN model was first applied to predict 52 solid-states and molecular properties, including formation energy, elastic constants, electronic bandstructure attributes, dielectric constants, and thermoelectric coefficients\cite{Choudhary2021_ALIGNN}. In almost every task, the ALIGNN model outperformed classical force-field inspired descriptors (CFID)\cite{choudhary2018machine} and the original CGCNN model\cite{Xie2018CGCNN} by yielding a lower mean absolute error (MAE) for predictions using comparable or improved training speed. In comparison to 18 other machine learning algorithms, the ALIGNN model also yields the lowest prediction MAE error for several tasks including band gap and formation energy prediction, as documented on the \texttt{matbench} website\cite{Dunn2020_matbench}. Since then, the ALIGNN model has also been used to guide new materials searches in the realm of metal-organic frameworks for carbon capture\cite{Choudhary2022_mof}, defect properties \cite{10.1063/5.0135382}, and high-$T_c$ conventional superconductors\cite{jarvis-supercond-bulk}. 

Since the initial presentation of the ALIGNN model in 2021\cite{Choudhary2021_ALIGNN}, additional models built from ALIGNN have been introduced. ALIGNN-d is an extension of the ALIGNN representation to explicitly include four-body dihedral angles, which was successfully used to predict the peak location and intensity in the optical spectra of Cu(II)-aqua complexes\cite{Hsu_ALIGNNd}. The de-ALIGNN model introduced by Gong \textit{et al.}\cite{Gong2022_deALIGNN} concatenates global descriptors of the material, such as average bond length or lattice parameters, to the learned features in the ALIGNN representation. Over 13 property prediction tasks, only two phonon-related tasks, phonon internal energy and heat capacity, showed large (>10 \%) prediction improvement in the de-ALIGNN model versus the original ALIGNN model. In Section \ref{sec:alignn_spec}, however, we show that phonon properties can be quickly and accurately predicted using ALIGNN through a direct prediction of the phonon density-of-states\cite{Gurunathan2023_ALIGNNphonons}.

In the following subsections, we will discuss specialized uses of ALIGNN to predict defect properties, spectral properties, and forces.

\subsubsection{ALIGNN-Spectra}\label{sec:alignn_spec} 

\begin{figure*}
    \begin{center}
        \includegraphics[width=0.8\textwidth]{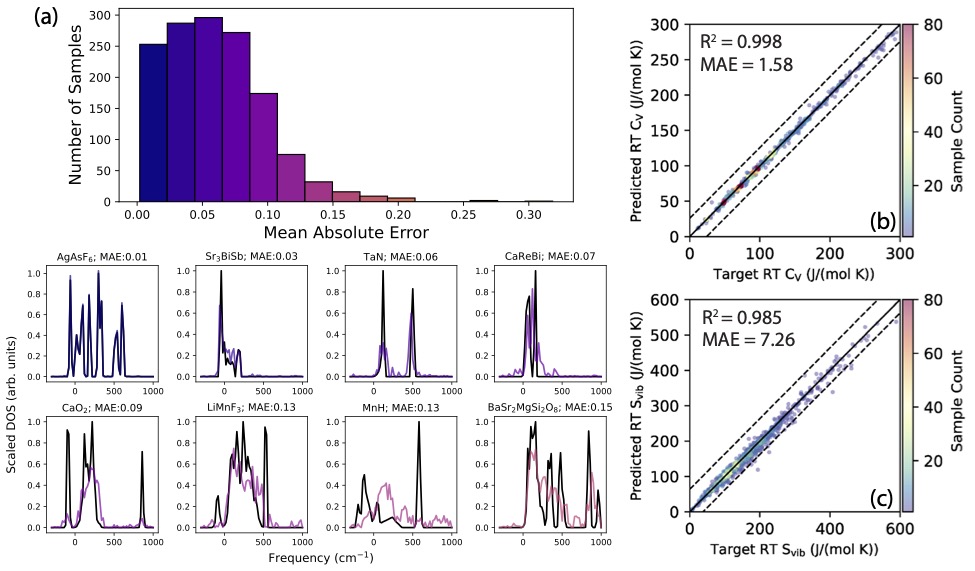}
        \caption{Performance of the ALIGNN phonon density-of-states (DOS) model. Panel (a) shows the distribution of ALIGNN-predicted phonon density of states (DOS) with respect to mean absolute error (MAE), indicating that 78 \% of samples show an MAE of less than 0.086. The example spectra below show the ALIGNN prediction (colored) against the DFT spectrum (black) to highlight the types of prediction errors that occur at each MAE level. In panels (b) and (c), we show that the room temperature heat capacity ($C_{\mrm{V}}$) and vibrational entropy ($S_{\mrm{vib}}$) derived from the ALIGNN phonon DOS closely corresponds to the target DFT-derived values. Reproduced with permission from Phys. Rev. Mater. 7, 023803 (2023). Copyright 2023 American Physical Society.}
        \label{fig:alignn_phonons}
    \end{center}
\end{figure*}

Thus far, the performance of ALIGNN has mainly been discussed in terms of scalar material property predictions, but the model has also been applied to predict spectral or frequency-dependent properties. The latter task requires multi-output predictions, which is relatively less well-developed for machine learning algorithms. Kaundiya \textit{et al.} first extended the ALIGNN model to enable multiple output features for the prediction of the electronic density-of-states (DOS)\cite{Kaundiya2022_ALIGNN_eDOS}. Two ALIGNN models trained using different representations of the DOS were compared: 1) a discretized electronic DOS with 300 evenly spaced frequency bins (D-ALIGNN), and 2) a low-dimensional representation of the electronic DOS generated using an autoenconder network with a latent dimensionality of 8, 12, 16, or 20 features (AE-ALIGNN). The D-ALIGNN model slightly outperformed the AE-ALIGNN model, but both yielded good prediction accuracy with over 80 \% of test samples showing an MAE of less than 0.2 states per eV per electron.

The phonon DOS is the subject of a later work, which further emphasizes DOS-derived properties obtained using a weighted integration of the phonon DOS\cite{Gurunathan2023_ALIGNNphonons}. These thermal and thermodynamic properties include the heat capacity ($C_{\mrm{V}}$), vibrational entropy ($S_{\mrm{vib}}$), and the phonon-isotope scattering rate ($\tau^{-1}_{\mrm{i}}$). The phonon DOS ALIGNN model was trained on a database of 14,000 DFT-computed phonon spectra calculated using the finite-difference method\cite{choudhary2020joint}. As shown in the histogram and example spectra of Figure \ref{fig:alignn_phonons}a, the spectra in the test set are concentrated at low prediction error levels and the ALIGNN model does a good job of capturing the location of peaks and general distribution of phonon modes, although the shape of the peaks is often altered. The ALIGNN phonon DOS predictions yield highly accurate estimates of the DOS-derived properties with correlation coefficients between ALIGNN and DFT values greater than 0.97 for all properties of interest. A general conclusion shown in this work is that a DOS-mediated approach outperforms a direct deep-learning approach for phononic properties. In other words, calculating properties like $C_{\mrm{V}}$ and $S_{\mrm{vib}}$ from the ALIGNN-predicted phonon DOS yields higher accuracy than training an ALIGNN model to predict $C_{\mrm{V}}$ or $S_{\mrm{vib}}$ directly. More details can be found in Ref. \onlinecite{Gurunathan2023_ALIGNNphonons}.

\subsubsection{ALIGNN-FF}\label{sec:alignn_ff}

In the previous sections, we described the application of the ALIGNN model for scalar and vector data, which are graph level outputs. Node level outputs such as forces, charges, and magnetic moments are the motivation for which the ALIGNN-atomwise model was developed. Specifically, for atomwise properties such as forces, they should be derivatives of energy and should be equivariant \cite{eqi-var} as the material system is rotated. For this task, we developed the ALIGNN force field (ALIGNN-FF) \cite{choudhary2023unified} to treat diverse crystals (chemically and structurally) with a combination of 89 periodic table elements. The entire JARVIS-DFT dataset was used to train ALIGNN-FF, which consists of 4 million energy-force entries (where 307113 are taken).

\begin{figure}
\begin{center}
\includegraphics[width=0.47\textwidth]{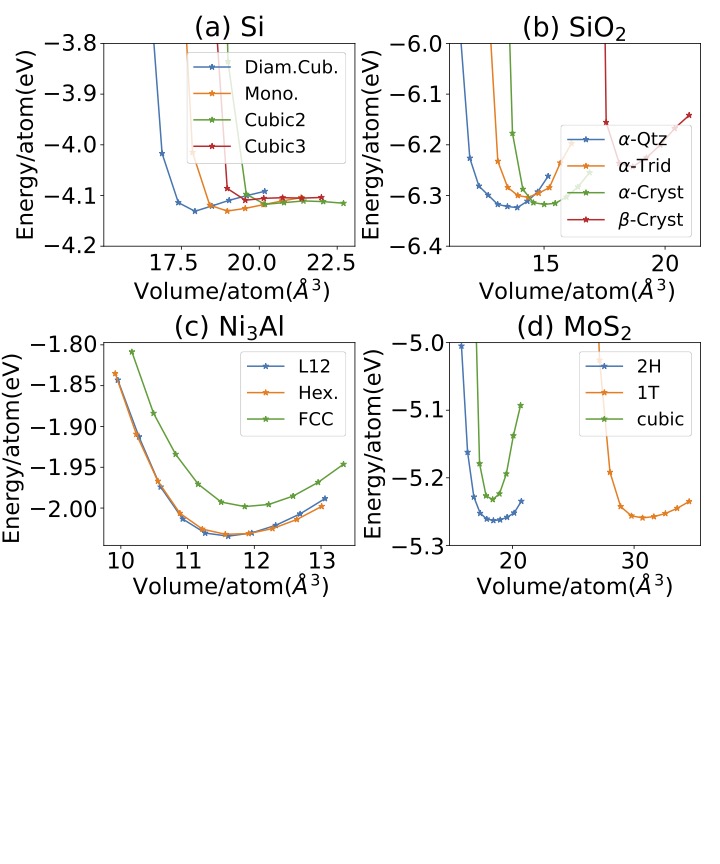}
\caption{ALIGNN-FF computed energy-volume curves for a) Si, b) SiO$_2$, c) Ni$_3$Al, d) MoS$_2$, with the ultimate goal of distinguishing polymorphs. Reproduced with permission from Digital Discovery, 2, 346-355 (2023). Copyright 2023 Royal Society of Chemistry.}
\label{fig:ALIGNNFF}
\end{center}
\end{figure}

Machine learning force fields (MLFFs) are a useful tool for the simulations of solids at a large scale. Previously, MLFFs have been designed for specific chemical environments, and are not usually able to be transferred to chemistries that differ from the training set. In recent years, efforts to develop a universal interatomic potential that is generalizable to diverse chemistries have been successful (i.e. M3GNET (MatGL) \cite{m3net} and GemNet-OC \cite{gasteiger2022gemnetoc}). 
We demonstrate ALIGNN-FF applicability beyond specific system-types, as it is built to predict atomistic properties for solids made of any combination of 89 periodic table elements. It was validated on predictions of properties such as lattice constants, and energy-volume curve. As an example, Fig. \ref{fig:ALIGNNFF} shows ALIGNN-FF ability to distinguish the polymorphs of various compounds (Si, SiO$_2$, Ni$_3$Al, and vdW-bonded MoS$_2$). 

ALIGNN-FF can also be used for fast optimization of atomic structures and structure prediction using evolution algorithms such as genetic algorithms. When compared to DFT and embedded-atom method (EAM) force fields, ALIGN-FF produced very similar equation of state curves. Moreover, ALIGNN-FF was used to optimize crystal structures in the crystallography open database (COD) as well as in the JARVIS database. For additional testing, ALIGNN-FF was used along with a genetic algorithm to predict the convex hull of a Ni-Al alloy system. Promisingly, the resulting convex hull reproduced the expected low energy structures without generating any unphysical low energy structures for Ni-Al phase diagram. As timing analysis shows that ALIGNN-FF is over 100 times faster than DFT methods, it can be used as pre-structure-optimizer before carrying out DFT calculations. More details on ALIGNN-FF can be found in Ref. \onlinecite{choudhary2023unified}. 


\subsubsection{ALIGNN-Superconducting}\label{sec:alignn_supercond}

\begin{figure}[h]
    \centering
    \includegraphics[trim={0. 0cm 0 0cm},clip,width=0.4\textwidth]{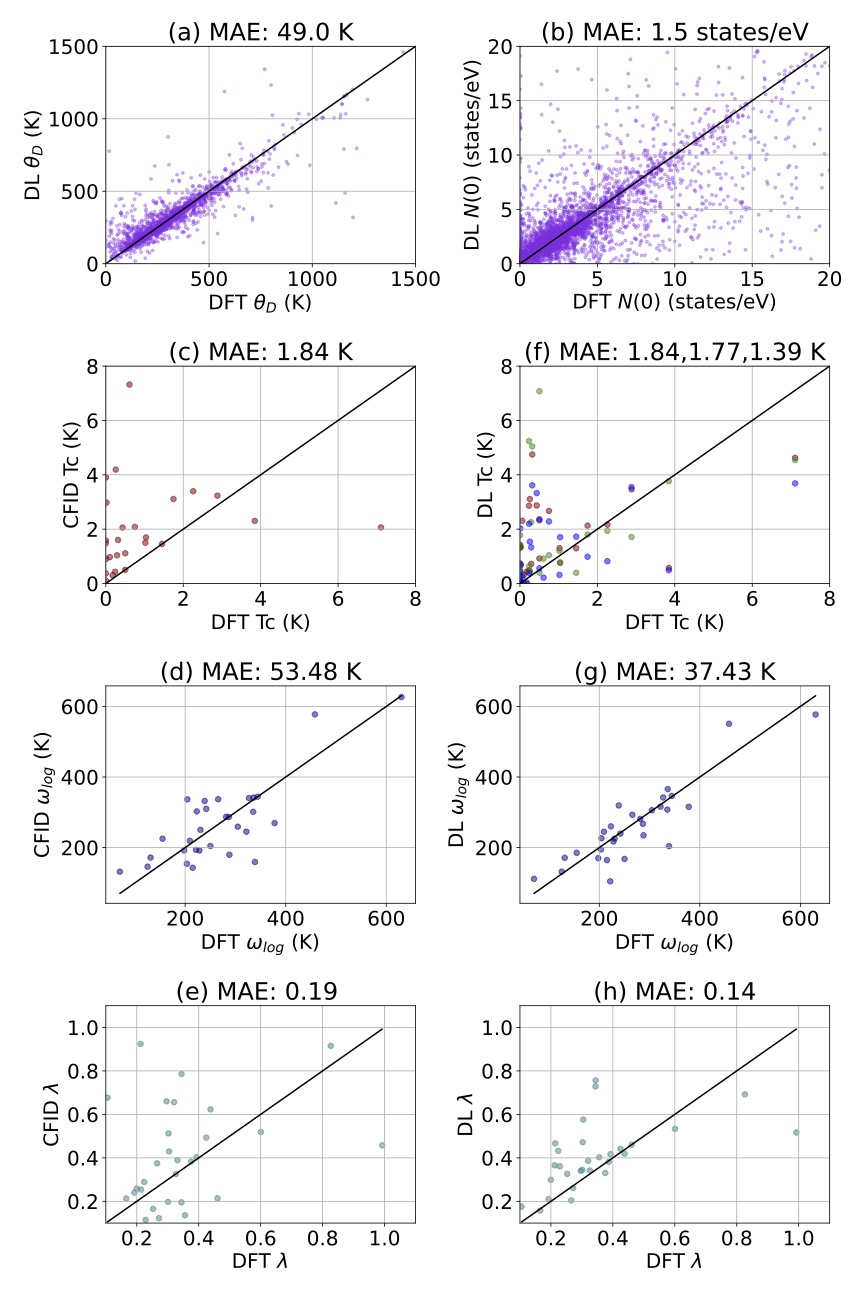}
    \caption{ALIGNN performance on a 5 \% test set for a) Debye temperature and b) electronic DOS. CFID (c,d,e) and ALIGNN (f,g,h) performance on a 5 \% test set for DFT computed $T_C$, $\omega_{log}$, and $\lambda$. Performance of a direct $T_C$ prediction (red), $T_C$ prediction using the Eliashberg function (black), and $T_C$ prediction utilizing the direct prediction of $\omega_{log}$ and $\lambda$ and then using McMillan-Allen-Dynes formula \cite{mcmillan1968transition} (green) are shown in (f). Reproduced with permission from npj Computational Materials 8, 244 (2022). Copyright 2022 Nature Publications.}
    \label{alignn1}
\end{figure}

In order to accelerate the initial BCS-inspired screening and direct computation of the electron-phonon coupling (EPC) parameters (see Section \ref{Bulk and 2D BCS Superconductors}), we developed deep learning tools (trained on JARVIS-DFT data) for direct property prediction from an arbitrary crystal structure. The BCS pre-screening step is far less computationally expensive than a full EPC calculation, but still requires a DFT calculation for the DOS at the Fermi level and $\theta_D$, which can still require significant computational expense. The results of these deep learning models for DOS and $\theta_D$ (on 5 \% held-out test sets) are depicted in Fig. \ref{alignn1}a) and b). 
Additionally, we developed machine-learning models to directly predict EPC properties trained exclusively on our DFT-PT calculations in Ref. \onlinecite{jarvis-supercond-bulk}. We used two methods (CFID and ALIGNN) and trained models for $T_C$ directly, in addition to training models for the EPC parameters ($\omega_{log}$ and $\lambda$). It is important to note that often deep learning models require much larger amounts of data. Although this is the case, our result show preliminary success with a smaller dataset, which is continually growing.

The performance of the CFID-based predictions are shown in Fig. \ref{alignn1}c), Fig. \ref{alignn1}d) and Fig. \ref{alignn1}e) for $T_C$, $\omega_{log}$ and $\lambda$ and the performance of the ALIGNN-based predictions are shown in Fig. \ref{alignn1}f, Fig. \ref{alignn1}g and Fig. \ref{alignn1}h.
It is clear that ALIGNN outperforms CFID in terms of computing $\omega_{log}$ while the MAE for $T_C$ and $\lambda$ are quite similar, indicating that it is easier to learn $\omega_{log}$. By computing $\omega_{log}$ and $\lambda$ with ALIGNN and plugging the quantities into the McMillan-Allen-Dynes equation \cite{mcmillan1968transition}, $T_C$ is predicted with an MAE of 1.77 K. Alternatively, we attempted to compute $T_C$ from the ALIGNN predicted Eliashberg function. We observed that the ALIGNN model can capture the peaks of the Eliashberg function. From this alternative method, we predict a $T_C$ with an MAE of 1.39, which has over a 24 \% improvement over the direct ALIGNN prediction. This implies that in comparison to direct ML predictions of properties, learning more fundamental physics-based quantities (such as the Eliashberg function) can be helpful for deep learning approaches with smaller amounts of data. Additionally, we used this superconducting ALIGNN model in conjunction with a generative diffusion model (crystal diffusion variational autoencoder \cite{xie2021crystal}) to inversely design new superconducting materials \cite{cdvae-sc}. More information on the ALIGNN superconducting model can be found in Ref. \onlinecite{jarvis-supercond-bulk}.  

\subsection{\label{sec:atomvision}AtomVision}
\begin{figure*}
\begin{center}
\includegraphics[width=0.8\textwidth]{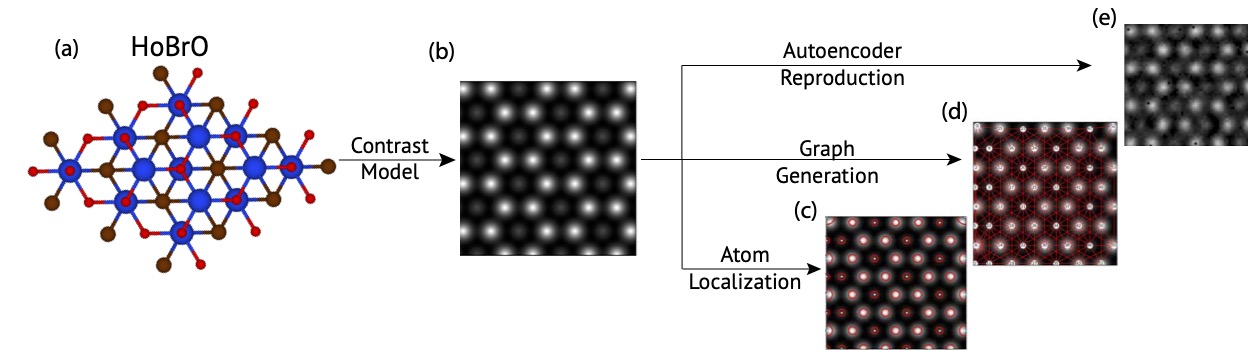}
\caption{Schematic of select capabilities in the AtomVision package. (a) An example 2D crystal structure is sampled from the JARVIS-2D DFT database\cite{choudhary2020joint} and a Rutherford scattering contrast model is applied to produce a synthetic HAADF-STEM image (b). Next, image analysis tasks implemented in the package are demonstrated, including: (c) localizing atom positions, (d) generating a non-Euclidean graph over the image, and (e) reconstructing the image from a low-dimensional representation using an autoencoder. Reproduced with permission from J. Chem. Inf. Model., 63, 6, 1708–1722 (2023). Copyright 2023 American Chemical Society.}
\label{fig:atomvision}
\end{center}
\end{figure*}

The AtomVision library is designed to be a general toolkit for both generating and analyzing image databases\cite{Choudhary2023_AtomVision}. Currently, the library implements contrast models that can be used to simulate scanning tunneling microscope (STM) and high angle annular dark field (HAADF) scanning transmission electron microscope (STEM) images given the crystal structure, and easily generate databases of simulated atomistic images. The STM images are computed using the Tersoff-Hamann formalism, which models the STM tip as an s-wave spherical state\cite{Tersoff_Hamann}. The HAADF STEM images are simulated using a convolution approximation often applied to thin film samples. This method convolves a point-spread function centered around the probe with a transmission function that considers the atomic number of the imaged specimen\cite{Combs2019_STEM}. The atomic number ($Z$) dependence of the intensity of the imaged atom is roughly proportional to $Z^2$, as predicted by Rutherford scattering\cite{Yamashita2018_Zcontrast}. Relevant images can also be curated from the literature through integration with the ChemNLP natural language processing package\cite{ChemNLP_arxiv} (more information in Section \ref{sec:chemnlp}). In contrast to other image datasets, which focus on a specific chemistry, the AtomVision package prioritizes chemical and structural diversity.

Numerous analysis tools are also provided, primarily based on machine learning methods. Although the datasets published with the package are focused on STM and STEM images, the analysis scripts can be used to easily train deep learning methods on any user-provided image data by simply providing directory paths for the training and test set of images.

First, the t-distributed stochastic neighbor embedding (t-SNE)\cite{vandermaaten08a_tSNE} is implemented, which performs a dimensionality reduction in the high-dimensional image data, allowing the spread of samples (images) to be visualized in a two or three-dimensional plot. The Euclidean distance between data points in a t-SNE plot relates to their similarity; however, the distances can only be interpreted qualitatively. Images that cluster together in the plot will tend to be more similar in their featurization, be it pixel intensity, red, green, blue triplets, or graph representations of images, which will be described in the next paragraph.

One critical image analysis task that is implemented in AtomVision, known as segmentation, consists of classifying pixels based on whether they compose the background or an object of interest. We utilize the U-Net pre-trained model\cite{ronneberger2015u} to distinguish atoms from background in the atomistic images. After this pixelwise classification, we can identify atom positions as well as characteristics of their intensity peak (e.g. peak width, maximum intensity) using a blob detection method implemented in the scikit-learn package\cite{scikit-learn}. With the atom positions identified, it is then possible to construct a non-Euclidean graph representation of the atomistic image. The atom peaks in the image become the nodes, which are featurized using the blob characteristics described above, and edges (representing bond vectors) are formed between atoms using the k-d tree nearest neighbor search algorithm. An additional line graph can then be constructed, as in the ALIGNN model, where bond vectors are the graph nodes and bond angles are the graph edges. 

There are then two main representations of images in the AtomVision package: 2D arrays of pixel intensities, and the graph representation. The AtomVision package allows the user to train neural networks based on either image representation to perform tasks such as image classification. The pixel-based data can be used to train convolutional neural networks (CNN) based on popular frameworks that include VGG\cite{Simonyan2014_VGG}, ResNet\cite{He2015_ResNet}, and DenseNet\cite{Huang2016_DenseNet} amongst others that are included in the AtomVision package. Similarly, the graph-based data can be used to train graph neural networks like the ALIGNN method described in Section \ref{sec:alignn}. We demonstrate both model types in an image classification task performed on the HAADF STEM image dataset in which we train a model to classify images into the 2D Bravais lattice type of the material in the image. In this particular test, the best performing CNN was the DenseNet with a classification accuracy of 83 \%, while the ALIGNN classifier had an accuracy of 78 \%. 

The pixelated images in the atomistic image datasets are described by 50,176 pixel intensities, and therefore exist in a very high-dimensional feature space. The manifold hypothesis suggests that the underlying structure of the data can be described by relatively fewer dimensions using feature extraction methods. The AtomVision package facilitates training and usage of an autoencoder network, which can create a low-dimensional representation of the image and then reconstruct the pixelated image from that latent representation. 

The final functionality currently implemented in the AtomVision package is a super-resolution generative adversarial network (SRGAN) model\cite{ledig2017photo}, which can upsample a low-resolution image to produce a high-resolution image. The SRGAN uses two separate deep learning models that compete with each other during optimization: 1) the generator produces a super-resolution image by first performing a feature extraction step, akin to the autoencoder model, and then interpolating within the learned latent representation; and 2) the discriminator should classify the image as either real or generated. As a result of the competition loop, the generator learns to ``trick" the discriminator model with increasingly realistic super-resolution images. We apply this model to the atomistic images and demonstrate a successful conversion of a low resolution image (64x64 pixels) to a high resolution image (256x256 pixels), showing the same image window. Additional information about the AtomVision package can be found in Ref. \onlinecite{Choudhary2023_AtomVision}.

\subsection{\label{sec:chemnlp}ChemNLP}

\begin{figure}
    \centering
    \includegraphics[trim={0. 0cm 0 0cm},clip,width=0.5\textwidth]{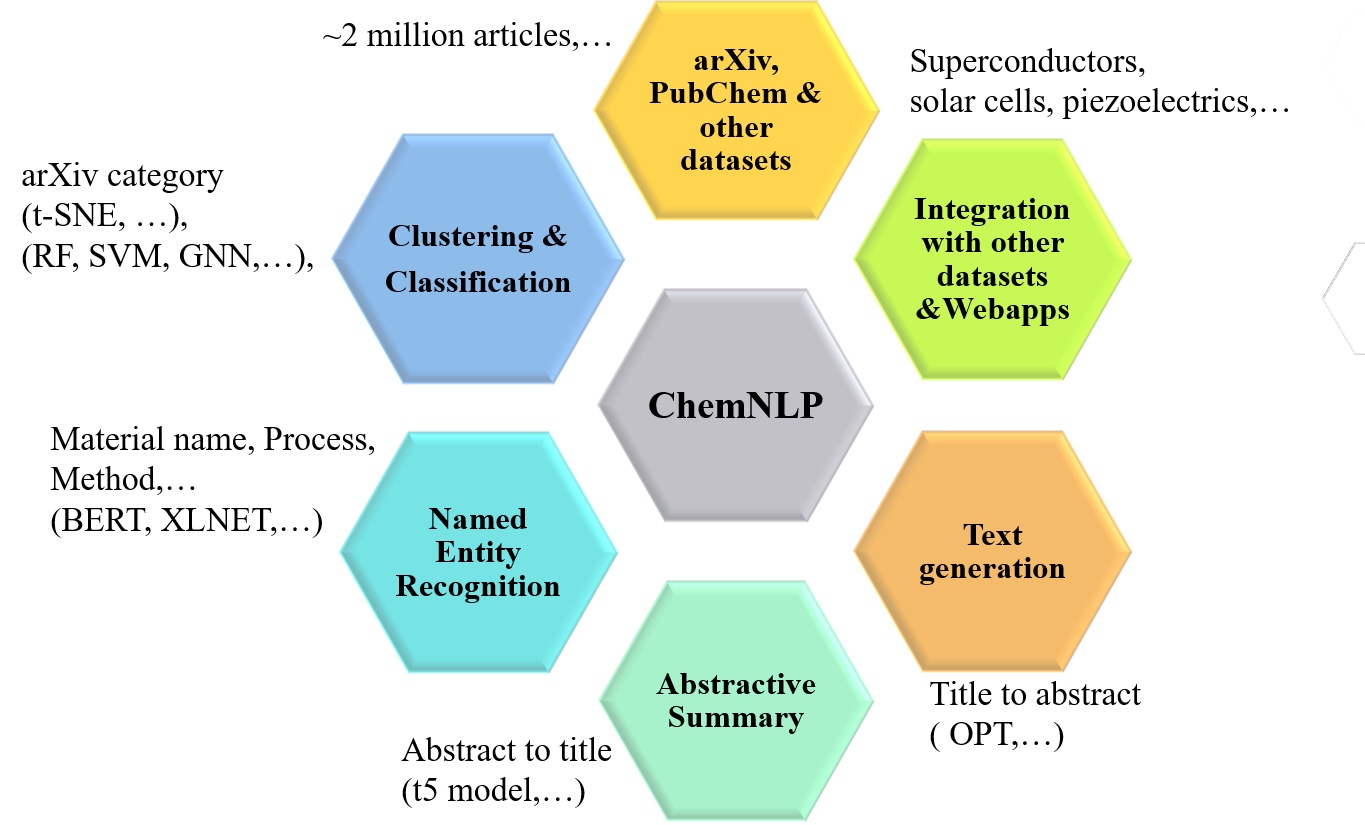}
    \caption{\label{fig:schematic}{A schematic overview of ChemNLP. The goal of ChemNLP is to provide a software toolkit with integrated dataset and comprehensive AI/ML tools for expanding natural language processing technique applications for tasks such as text classification, clustering, named entity recognition, abstractive summarization and text generation.}}
\end{figure}

Much of the data on materials science is available in text format in the form of articles that are not easily amenable to standard automated analysis. To address this barrier, we developed ChemNLP \cite{ChemNLP_arxiv},  a library that utilizes natural language processing (NLP) for chemistry and materials science data. Currently, ChemNLP is based on publicly available platforms such arXiv (\url{https://arxiv.org/}), Pubchem (\url{https://pubchem.ncbi.nlm.nih.gov/}) datasets and Huggingface (\url{https://huggingface.co/}) \cite{wolf2020huggingfaces} libraries. 

ChemNLP organizes the NLP data and tools for materials chemistry application in a format suitable for model training. In addition to data curation, it allows the integration of useful analyses such as: 1)  classifying and clustering texts based on their categories, 2) named entity recognition for large-scale text-mining, 3) abstractive summarization for generating titles of articles from abstracts, 4) text generation for suggesting abstracts from titles, 5) integration with density functional theory datasets for identification of potential candidate materials, and 6) web-interface development for text and reference query. A schematic of the ChemNLP library is given in Fig. \ref{fig:schematic}. 

ChemNLP uses several conventional machine learning algorithms as well as state of the art transformer models for comparison and validation. Some of the algorithms in ChemNLP include support vector machines, random forest, graph neural networks, Google's T5 \cite{raffel2020exploring}, OpenAI's GPT-2 \cite{brown2020language} and Meta AI's OPT \cite{zhang2022opt} transformer models, all of which are fine-tuned on materials chemistry text data.
The web-app of ChemNLP allows for the searching of various text information (such as material properties, synthesis procedure, etc.) given the chemistry information (stoichiometry). As an application of transformer models, ChemNLP showed that fine-tuning general large language models (LLMs) for abstract to title and vice-versa can result in an improvement in performance compared to the original pre-trained model.

Specifically, we applied text classification models for the arXiv and PubChem datasets where we chose title, abstracts, and titles along with abstracts to classify the articles. We used ML algorithms such as random forest, support vector machine, logistic regression, and graph neural network and found that the highest classification accuracy (91.3 $\%$ for arXiv and 97.6 $\%$ for PubChem) was achieved for linear support vector machine. In order to mine text and extract meaningful information, named entity recognition or token classification can be used. Information such as material name, sample descriptor, symmetry label, synthesis/characterization method, property and application can be extracted and utilized. We used the MatScholar \cite{matscholar} dataset to train a transformer model with XLNet \cite{yang2020xlnet} and applied it to the arXiv titles and abstracts and full texts, where we found the F1 score to be 87 $\%$. With regards to text-to-text generation models, we focused on abstract summarization (creating a title from the abstract) and text generation (generating an abstract from the title). We used a pre-trained T5 \cite{raffel2020exploring} model for abstract summarization and fine-tuned it for the arXiv dataset. As a metric of success, we used the ROGUE (Recall-Oriented Understudy for Gisting Evaluation) score. For the fine-tuned T5 model, we obtained a ROGUE score of 46.5 $\%$ (as opposed to the 30.8 $\%$ of the pre-trained model). To generate the abstract from the title, we fine-tuned a pre-trained OPT \cite{zhang2022opt} model and obtained a ROGUE score of 37 $\%$. This proves that ChemNLP provides a flexible format to fine-tune existing text generation models that may be developed in the future.

Finally, ChemNLP allows a seamless integration of arXiv and DFT databases (i.e. JARVIS-DFT). In our work, we demonstrated how ChemNLP can aid in the discovery of new superconductors by simultaneously searching the JARVIS Superconductor dataset \cite{jarvis-supercond-bulk} and the arXiv dataset. With regards to materials with a $T_c$ above 1 K, we found 635 in the DFT dataset and 1071 chemical formulas in the arXiv dataset, with only 43 common materials. This integration of literature and calculated results can motivate further screening of potential candidate superconductors. In addition to aiding the search for materials for specific applications, ChemNLP can be used along with DFT databases to generate formatted descriptions of atomic structure information that can be used for training future large language models (i.e. json formatted text). Further details of ChemNLP and the success metrics used (classification accuracy, F1 score, ROGUE score) can be found in Ref. \onlinecite{ChemNLP_arxiv}.

\subsection{\label{sec:UQ}Uncertainty Analysis}

Uncertainty quantification in ML-based material property prediction is important for assessing the accuracy and reliability of machine learning methods for material property predictions \cite{doi:10.1146/annurev-matsci-070218-010015,https://doi.org/10.1002/adma.202104113}. For example, if the uncertainty in the prediction is not known or is too large, predictions can be challenged. For this reason, the field of uncertainty quantification for materials AI/ML-based predictions is a field which could benefit from advancements \cite{doi:10.1146/annurev-matsci-070218-010015,https://doi.org/10.1002/adma.202104113}. Confidence intervals are widely reported for ML predictions, but the evaluation of individual uncertainties on each prediction (prediction intervals) are not as commonly reported. Due to this, JARVIS has focused on individual property uncertainty predictions for ML models \cite{uncertainty-ml}. To compute individual uncertainties, we specifically used machine learning the prediction intervals directly, Quantile loss function, and Gaussian processes. These uncertainty prediction methods were tested and compared for 12 ML-computed properties. The JARVIS-DFT dataset was used for all training and testing.

In summary, we found that direct modeling of the individual uncertainty is favored due to the fact that the overestimation and underestimation of the errors is minimized in most cases. In addition, it is the easiest method to fit and implement \cite{uncertainty-ml}. The Quantile method requires the fitting of three different models. Gaussian processes give a reasonable estimate for the prediction intervals, but are overestimated and more time consuming to fit compared to the other methods. Additionally, direct prediction of individual uncertainty has an advantage because it allows the use of any loss function. One caveat is that it requires splitting the data into three parts, which can be a potential issue if the dataset is too small. The codes developed for evaluating the prediction intervals are publicly available within JARVIS-tools. More details of this work can be found in Ref. \onlinecite{uncertainty-ml}. 




\section{\label{sec:qc}Quantum Computation}

\begin{figure*}
\begin{center}
\includegraphics[width=0.8\textwidth]{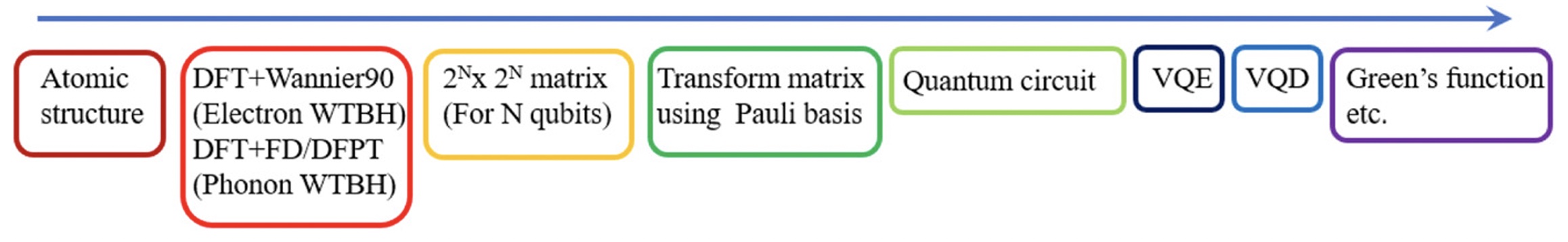}
\caption{The steps used in predicting phonon and electron properties of a material on a quantum computer. Reproduced from with permission from J. of Phys. Cond. Matter 33, 385501 (2021). Copyright 2021 IOP Science Publishing.}
\label{fig:atomqc}
\end{center}
\end{figure*}

The ability to solve quantum chemistry problems is one of the most promising near-term applications of a quantum computer \cite{nielsen2001quantum}. Variational quantum eigen solver (VQE) \cite{vqe} and variational quantum deflation (VQD) \cite{Higgott2019variationalquantum} are Quantum algorithms that have been applied to molecules \cite{vqe}. There is a strong desire to implement these Quantum algorithms for crystals. As a result, we have developed the AtomQC \cite{ChoudharyAtomQC} package, which adds quantum computation tools to the JARVIS infrastructure.

WTBH approaches were utilized to demonstrate the application of VQE and VQD to compute electronic and phonon properties of various materials, including elemental solids and and multi-component systems. For 307 spin–orbit-based electronic WTBHs and 933 finite-difference-based phonon WTBHs, we applied VQE and VQD algorithms. We only deal with the single-particle picture in this study, but we strongly believe our work can set the course for the solution of interacting Hamiltonians. Such interacting Hamiltonians can be obtained from methods such as dynamical mean-field theory (DMFT) and Green’s function with screened Coulomb potential theory (GW), which can be a much more suitable problem to simulate on a quantum computer. A preliminary workflow that combines the VQD algorithm with DMFT-based solving of the lattice Green’s function is provided in this work. Fig. \ref{fig:atomqc} shows the entire quantum computation workflow. These WTBH solvers can be used to test other various Quantum algorithms and are publicly available. More information on AtomQC can be found in Ref. \onlinecite{ChoudharyAtomQC}.

\section{\label{sec:exp}Experiments}

Although, majority of the data in JARVIS is originated from computation, we use data validation and benchmarking with experiments whenever applicable. In some cases, we obtain experimental data from literature or from in-house standard reference material (SRM) data at NIST. In other cases, we perform our own experiments along with our computational efforts. This  experimental data includes XRD and neutron diffraction patterns, CO$_2$ adsorption isotherms \cite{co2-rr}, magnetic susceptibility measurements \cite{2dsc}, spectroscopic ellipsometry dielectric functions, Raman spectra, STM/STEM images and transport measurements.

\begin{figure}
\caption{
Experimental DC magnetic susceptibility as a function of temperature (used to determine $T_c$) for layered: a) 2H NbSe$_2$, b) 2H-NbS$_2$, c) FeSe, and d) ZrSiS. Reproduced with permission from Nano Letters 23, 969-978 (2023). Copyright 2023 American Chemical Society.}
\begin{center}
\includegraphics[width=8cm]{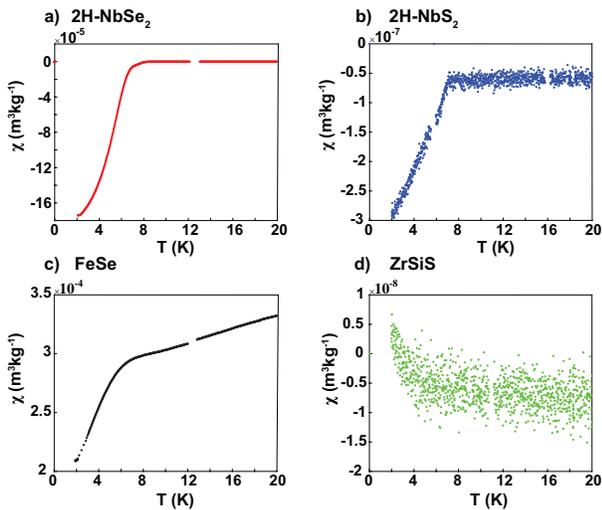}
\label{exp}
\end{center}
\end{figure}

Most recently and notably, we have conducted our own experiments for magnetic topological materials and 2D superconductors. With regards to magnetic toplogical materials, we measured the Anomalous Hall effect of CoNb$_3$S$_6$ and conducted inverse spin-Hall signal measurements for Mn$_3$Ge, two materials theoretically predicted to be topological materials in Ref. \onlinecite{PhysRevB.103.155131} (see Section \ref{topscreen}). We performed zero-field-cooled magnetometry experiments to determine the critical temperature of selected 2D superconductors to verify the theoretical predictions in Ref. \onlinecite{2dsc} (see Section \ref{Bulk and 2D BCS Superconductors}). We conducted these experiments for layered 2H-NbSe$_2$, 2H-NbS$_2$, FeSe, and ZrSiS. Fig. \ref{exp} depicts the measured magnetic susceptibility (using a magnetic field strength of 0.01 T) as a function of temperature. We observe that out of these layered materials, 2H-NbSe$_2$ has a $T_c$ of 8.3 K, 2H-NbS$_2$ has a $T_c$ of 7.1 K, FeSe has a $T_c$ of 7.5 K, and ZrSiS does not have a superconducting transition due to the measured decreasing magnetic susceptibility with increasing temperature. We discuss these measurements within the context of our DFT-computed results in Ref. \onlinecite{2dsc}.

These experimental datasets are now being integrated in the JARVIS-Leaderboard for benchmarking and validation purposes (see Section \ref{leaderboard}). Some simulated experimental data, such as XRD patterns,
can be computed with JARVIS-Tools. As the experimental
datasets in JARVIS are not exceedingly large at the moment, we can currently apply machine learning algorithms on computational data, and in the future, apply the same pipeline to experimental data.

\section{\label{sec:opt}Optimade and Nomad}

Data in JARVIS are being integrated within large-scale data efforts such as NOMAD \cite{draxl2018nomad} and OPTIMADE \cite{andersen2021optimade} for sustainability, and inter-operatability.

The Open Databases Integration for Materials Design (OPTIMADE) consortium has designed a universal application programming interface (API) to make materials databases accessible and interoperable. The OPTIMADE API has a set of well defined key-value pairs such as chemical formula name, number of elements etc. for each atomic structure which allows sending a universal API search for multiple data-efforts. The implementation required a django-rest-framework integration with specific data-models, pagination and other specification as detailed in OPTIMADE to be compatible with other infrastructures.

Similarly, NOMAD project allows storage of raw data files and provides several interactive GUI tools. Although, JARVIS has its own storage mechanism, having data distributed on platforms such as NOMAD and OPTIMADE allows enhanced transparency, which is essential for large scale data-driven materials design.




\section{\label{sec:notebooks}JARVIS Tutorial Notebooks}

A collection of interactive \textsc{python} notebooks are hosted in the jarvis-tools-notebooks GitHub Repository\cite{JarvisNbGitHub} that can be run on a user's local computer or easily through a cloud-based \textsc{python} development environment. All package installation steps are included in the notebooks such that they can be executed and edited in a standalone fashion, allowing users to easily make use of JARVIS models. These interactive notebooks are meant to supply an example calculation that will execute quickly and reproduce a portion of the results in a JARVIS-associated publication. The collection of notebooks include machine learning models that allow users to train and utilize ALIGNN or AtomVision models for material property prediction and image classification. There are also several electronic structure and atomistic calculation notebooks that analyze DFT, tight-binding, or MD outputs to calculate material properties including elastic properties, spin-orbit spillage, dielectric functions, thermoelectric and photovoltaic properties. Finally, a notebook based on the AtomQC and qiskit\cite{Qiskit} packages for quantum computation is provided.

\section{\label{sec:lb}Leaderboard}\label{leaderboard}
The JARVIS-Leaderboard \cite{choudhary2023large} is a large-scale benchmarking effort for various computational and experimental methodologies for materials science applications. The main goal of the JARVIS-Leaderboard is to enhance reproducibility and transparency for various methodologies within the materials science field. Although leaderboard efforts have previously been developed for specific applications (i.e. MatBench \cite{dunn2020benchmarking}, OpenCatalystProject \cite{chanussot2021open}, etc.), there lacks a benchmarking platform with multiple data modalities for perfect and defective materials as well as ease to add new benchmarks. The JARVIS-Leaderboard (\url{https://pages.nist.gov/jarvis_leaderboard/}) attempts to bridge the gap between different methods and material classes by allowing users to set up benchmarks and make contributions in the form of datasets, codes and meta-data submissions (through GitHub actions). These contributions are compared with experimental data where applicable and the accuracy of each contribution is assigned a ``score" (MAE with respect to the ground truth). Some of the categories are: Artificial Intelligence (AI), Electronic Structure (ES), Quantum Computation (QC) and Experiments (EXP).

For AI, various methods (descriptor-based, neural network-based) and data (atomic structure, atomistic images, spectra, text) are benchmarked. For ES, multiple approaches (DFT, QMC, Tight binding, GW), software
packages, and pseudopotentials are considered and the results are compared to experiments whenever applicable. Multiple FF approaches for material property predictions are compared (classical FF, MLFF). For QC, we compare the performance of various quantum algorithms and circuits for Hamiltonian simulations. For experiments, inter-laboratory (round robin) approaches are used. In addition to prediction results, we attempt to capture the underlying software, hardware and instrumental frameworks to enhance reproducibility and method validation, which can aid in developing new and more reliable techniques. Currently there are over 1400 user contributions using over 150 different methods, and these numbers are growing rapidly.  

\section{\label{sec:outreach}External Outreach}

In addition to the databases, tools and applications that are part of the JARVIS infrastructure, there has been substantial effort devoted to outreach in the materials science research community. The JARVIS team has annually hosted the Artificial Intelligence for Materials Science (AIMS) and Quantum Matters in Materials Science (QMMS) workshops, where speakers have been invited from academia, government and industry to discuss key achievements and challenges in the respective fields. Topics presented at AIMS include: dataset and tools for employing AI for materials, integrating experiments with AI techniques, graph neural networks, comparison of AI techniques for materials, the challenges of applying AI to materials, uncertainty quantification and building trust in AI predictions, generative modeling, using AI to develop classical force-fields, natural language processing and AI-guided autonomous experimentation. Topics presented at QMMS include: discovery and characterization of new materials; optimization of known quantum materials, investigation of defect induced behavior and transitions;  electronics, spintronics, quantum memory applications, challenges in applying quantum information systems technologies at industrial scale, and accurate many-body computational methods to treat quantum materials. 

The JARVIS team has also organized a series of hands-on workshops at different academic and government institutions, known as JARVIS-Schools. JARVIS-Schools consist of a tutorial and hands-on session to introduce open-access databases and tools for materials-design. These sessions are accompanied by a series of power-point presentations on the core-topics, Google-Collab/Jupyter notebook examples and discussion. The hands-on session/discussion topics include: electronic structure calculations (DFT, tight-binding etc.), density functional theory for predicting properties (e.g., solid-state materials), machine learning (for atomistic, image and text data), quantum computation and its applications to materials, and classical force-field calculations for large scale properties. An updated calendar of JARVIS events can be found here: \url{https://jarvis.nist.gov/events/}. 

There are various mechanisms to collect external usage data for JARVIS inside and outside the materials science community. These sources include the number of users registered for the JARVIS API, Google analytics results for viewers, number of citations for papers, downloads of software tools on GitHub, number of views and downloads from the Figshare repository, number of attendees in the AIMS/QMMS workshops and JARVIS-Schools and number of collaborators developed inside and outside NIST from academia, national labs and industry. A wide usage of JARVIS resulted in JARVIS being highlighted as a standard platform for materials design in the NIST US CHIPS Acts strategic plan (\url{https://www.nist.gov/chips/implementation-strategy}). 



\section{Data Availability and Software}
The data and software mentioned in this review article are available at \url{https://jarvis.nist.gov/} and \url{https://github.com/usnistgov/jarvis}. JARVIS-Tools is an open-access software package for atomistic data-driven materials design. JARVIS-Tools can be used for a) setting up calculations, b) analysis and informatics, c) plotting, d) database development and e) web-page development. Software used in workflow tasks for pre-processing, executing, and post-processing include: VASP \cite{kresse1996efficient,kresse1996efficiency}, Quantum Espresso \cite{giannozzi2009quantum,giannozzi2020quantum}, Wien2k \cite{doi:10.1063/1.5143061}, BoltzTrap \cite{MADSEN200667}, Wannier90 \cite{MOSTOFI20142309}, QMCPACK \cite{Kim_2018,doi:10.1063/5.0004860}, LAMMPS \cite{LAMMPS}, scikit-learn \cite{scikit-learn}, TensorFlow \cite{tensorflow2015-whitepaper}, LightGBM \cite{ke2017lightgbm}, Qiskit \cite{Qiskit}, Tequila \cite{Kottmann_2021}, Pennylane \cite{bergholm2022pennylane,arrazola2023differentiable}, Deep Graph Library \cite{wang2020deep}, PyTorch \cite{paszke2019pytorch}. JARVIS databases such as JARVIS-DFT, FF, ML, WannierTB, Solar, and STM can be downloaded. Raw input and output files can be accessed from JARVIS databases to enhance reproducibility in calculations. Different descriptors, graphs and datasets for training machine learning models are also included in JARVIS-Tools. Capabilities can be easily be extended to HPC systems (Torque/PBS and SLURM). Documentation for JARVIS-Tools, including installation instructions, can be found here: \url{https://pages.nist.gov/jarvis/}.

 \section{Notes}
Certain commercial equipment or materials are identified in this paper to adequately specify the experimental procedures. In no case does the identification imply recommendation or endorsement by NIST, nor does it imply that the materials or equipment identified are necessarily the best available for the purpose. The authors declare no competing interests. Please note that the use of commercial software (VASP) does not imply recommendation by the National Institute of Standards and Technology.

\section{Acknowledgments}
All authors thank the National Institute of Standards and Technology for funding, computational, and data-management resources. K.C. thanks the computational support from XSEDE (Extreme Science and Engineering Discovery Environment) computational resources under allocation number TG-DMR 190095. Contributions from K.C. were supported by the financial assistance award 70NANB19H117 from the U.S. Department of Commerce, National Institute of Standards and Technology.

\section*{References}

\bibliography{jarvisbib}

\end{document}